\documentclass[twocolumn]{aastex631}

\usepackage{graphicx}	\usepackage{amsmath}	\usepackage{amssymb}	\usepackage{hyperref}
\usepackage{xcolor}     

\usepackage{graphicx,color}	\usepackage{cancel}

\newcommand{\kms}{\textrm{km s}^{-1}}
\newcommand{\hi}{\text{H\,\sc{i}}}
\newcommand{\Msol}{\textrm{M}_{\odot}}

\newcommand{\barolo}{\textsc{3DBAROLO}}
\newcommand{\fat}{\textsc{FAT}}
\newcommand{\wkapp}{\textsc{WKAPP}}
\newcommand{\NA}[1]{\textcolor{black}{#1}}

\newcommand{\queens}{Department of Physics, Engineering Physics, and Astronomy,Queen's University,\\ Kingston ON K7L~3N6, Canada}
\newcommand{\ICRAR}{International Centre for Radio Astronomy Research (ICRAR), The University of Western Australia,\\ 35 Stirling Highway, Crawley WA 6009, Australia}
\newcommand{\ASTRO}{ARC Centre of Excellence for All Sky Astrophysics in 3~Dimensions (ASTRO~3D),\\ Australia}
\newcommand{\CSIRO}{CSIRO Space and Astronomy, PO Box 1130, Bentley WA 6102, Australia}
\newcommand{\ICRARCURTIN}{International Centre for Radio Astronomy Research (ICRAR) - Curtin University, \\Bentley, WA 6102}

\begin{document}

\title{WALLABY Pilot Survey: Gas-Rich Galaxy Scaling Relations from Marginally-Resolved Kinematic Models}

\correspondingauthor{N. Deg}
\email{nathan.deg@queensu.ca}

\author[0000-0003-3523-7633]{N. Deg}
\affiliation{\queens}

\author[0000-0002-3929-9316]{N. Arora}
\affiliation{\queens}

\author{K. Spekkens}
\affiliation{Department of Physics and Space Science, Royal Military College of Canada,\\ P.O.\ Box 17000, Station Forces Kingston ON K7K~7B4, Canada}
\affiliation{\queens}

\author{R. Halloran}
\affiliation{\queens}

\author{B.  Catinella}
\affiliation{\ICRAR}
\affiliation{\ASTRO}

\author[0000-0002-5434-4904]{M.~G.~Jones}
\affiliation{Steward Observatory, University of Arizona, 933 North Cherry Avenue, Rm. N204, Tucson, AZ 85721-0065, USA}

\author[0000-0003-0509-1776]{H. Courtois}
\affiliation{University Claude Bernard Lyon 1, IUF, IP2I Lyon, 69622, Villeurbanne, France}

\author[0000-0002-3254-9044]{K. Glazebrook}
\affiliation{Centre for Astrophysics \& Supercomputing, Swinburne University, Hawthorn, VIC 3122, Australia}

\author{A. Bosma}
\affiliation{Aix Marseille Univ, CNRS, CNES, LAM, Marseille}

\author{L. Cortese}
\affiliation{\ICRAR}
\affiliation{\ASTRO}

\author{H. D\'{e}nes}
\affiliation{School of Physical Sciences and Nanotechnology, Yachay Tech University, Hacienda San Jos\'{e} S/N, 100119, Urcuquí, Ecuador }

\author{A. Elagali}
\affiliation{Department of Biological sciences, University of Western Australia }

\author{B.-Q. For}
\affiliation{\ICRAR}
\affiliation{\ASTRO}

\author{P. Kamphuis}
\affiliation{Ruhr University Bochum, Faculty of Physics and Astronomy, Astronomical Institute (AIRUB), 44780 Bochum, Germany}

\author{B.S. Koribalski}
\affiliation{Australia Telescope National Facility, CSIRO, Space and Astronomy, P.O. Box 76, Epping, NSW 1710, Australia}
\affiliation{School of Science, Western Sydney University, Locked Bag 1797, Penrith, NSW 2751, Australia}

\author{K. Lee-Waddell}
\affiliation{\ICRAR}
\affiliation{\CSIRO}
\affiliation{\ICRARCURTIN}

\author[0000-0001-5175-939X]{P. E. Mancera Pi\~na}
\affiliation{Leiden Observatory, Leiden University, P.O. Box 9513, 2300 RA, Leiden, The Netherlands}

\author{J. Mould}
\affiliation{Centre for Astrophysics \& Supercomputing, Swinburne University, Hawthorn, VIC 3122, Australia}
\affiliation{ARC Centre of Excellence for Dark Matter Particle Physics}

\author{J. Rhee}
\affiliation{\ICRAR}

\author{L. Shao}
\affiliation{National Astronomical Observatories, Chinese Academy of Sciences, Beijing 100101, China}

\author{L. Staveley-Smith}
\affiliation{\ICRAR}
\affiliation{\ASTRO}

\author{J. Wang}
\affiliation{Kavli Institute for Astronomy and Astrophysics, Peking University, Beijing 100871, China}

\author{T. Westmeier}
\affiliation{\ICRAR}
\affiliation{\ASTRO}

\author{O.I. Wong}
\affiliation{\CSIRO}
\affiliation{\ICRAR}
\affiliation{\ASTRO}

\begin{abstract}
We present the first set of galaxy scaling relations derived from kinematic models of the Widefield ASKAP L-band Legacy All-sky Blind surveY (WALLABY) pilot phase observations. Combining the results of the first and second pilot data releases, there are 236 available kinematic models. We develop a framework for robustly measuring \textcolor{black}{\hi\ disk structural properties} from these kinematic models; applicable to the full WALLABY survey.  Utilizing this framework, we obtained the \hi\ size, a measure of the rotational velocity, and angular momentum for 148 galaxies.  These comprise the largest sample of galaxy properties from an untargetted, uniformly observed and modelled \hi\ survey to date.  We study the neutral atomic Hydrogen (\hi) size-mass, size-velocity, mass-velocity, and angular momentum-mass scaling relations. We calculate the slope, intercept, and scatter for these scaling relations and find that they are similar to those obtained from other \hi\ surveys.  We also obtain stellar masses for 92 of the 148 robustly measured galaxies using multiband photometry through the Dark Energy Sky Instrument Legacy Imaging Survey Data Release-10 images.  We use a subset of 61 of these galaxies that have consistent optical and kinematic inclinations to examine the stellar and baryonic Tully Fisher relations, the gas fraction-disk stability and gas fraction-baryonic mass relations. These measurements and relations demonstrate the unprecedented resource that WALLABY will represent for resolved galaxy scaling relations in \hi.
\end{abstract}


\section{Introduction}

Galaxies present a wide spectrum of morphologies, sizes, and compositions \citep{Hubble1926,deVaucouleurs1959,Buta2013}.  Despite this diversity, galaxies show remarkable similarity in their underlying physical properties, as evidenced by the many observed scaling relations showing correlations between various galaxy properties, \textcolor{black}{such as size, mass, and rotation velocity in disk galaxies}.  Such scaling relations point to physical laws governing galaxy formation and evolution.  Given their importance, galaxy scaling relations have been studied in detail for many years \citep[][to name a few]{Faber1976,Tully1977,Fall1980,Broeils1997,Courteau2007,Hall2012,Kormendy2013,Stone2021,Arora2023}.

There are many different scaling relations \textcolor{black}{for disk galaxies,} covering many different wavelengths.  They include the various Tully Fisher Relations (TFR; \citealt{Tully1977,Mcgaugh2000, Verheijen2001, Hall2012, Lelli2016,Papastergis2016, Ponomareva2017,Lelli2019,Geha2016,McQuinn2022,Arora2023}), the \hi\ size-mass relation \citep{Broeils1997,Verheijen2001,Swaters2002,Noordermeer2005,Begum2008,Wang2013}, the size-velocity relation \citep{Courteau2007,Dutton2007, Avila-Reese2008, Saintonge2011, Ouellette2017, Lapi2018, Meurer2018,Stone2021, Arora2023}, the specific angular momentum-mass relation \citep{Fall1980,Romanowsky2012,Obreschkow2014,Murugeshan2020,Kurapati2021,ManceraPina2021a,Elson2023,Sorgho2024}, and more \citep[see Table 6 from][for a summary of optical scaling relations]{Stone2021}.  These different relations have provided a great deal of insight into galaxy structure and evolution.  Measuring their slope and intercept constrains the relation between various physical parameters, while the scatter often highlights other processes that affect a galaxy.

Measuring scaling relations requires high quality observations with well understood uncertainties.  It also requires \textcolor{black}{sufficiently large sample sizes} to obtain statistically meaningful \textcolor{black}{measurements}.  Ideally, these galaxies would come from a single survey with a known selection function. \textcolor{black}{This enables straightforward extractions of physical relations as well as quantitative comparisons between the survey and simulations.  However, it is also possible} to combine observations from various surveys together to obtain a large enough sample for the science being investigated.  \textcolor{black}{But combining different surveys} introduces additional complications to the analysis \textcolor{black}{due to differing systematics in the combined sample}.

\textcolor{black}{In \hi\ this challenge is heightened due to the low resolution.  Most scaling relations, such as the size-mass relation \citep{Wang2016}, require resolved observations, but the large \hi\ surveys \textcolor{black}{that have observed the \hi\ content of the largest number of galaxies} \citep{Meyer2004,Giovanelli2005, Haynes2018}, are single dish observations that do not resolve most of their detections.  However, the Widefield ASKAP L-band Legacy All-sky Blind surveY (WALLABY; \citealt{Koribalski2020}) on the Australian Square Kilometer Array Pathfinder (ASKAP) is changing this situation.}

WALLABY is an untargetted, nearly all-Southern sky \hi\ survey that will detect $\sim2\times10^{5}$ galaxies of which $\sim10^{4}$ will be resolved by 2-3+ resolution elements.  \textcolor{black}{This large sample makes WALLABY the premier survey to study the population-wide statistics of gaseous scaling relations.}  WALLABY observations are underway with the first pilot data release (PDR1; \citealt{Westmeier2022}) already public; containing $\sim600$ galaxies and $\sim110$ kinematic models \citep{Deg2022}.  A second pilot data release (PDR2; \citealt{Murugeshan2024}) will contain an additional $\sim1800$ galaxies and $\sim 130$ kinematic models.  Taken together, the WALLABY pilot data releases (PDR) already have produced sample of $236$ uniformly observed resolved galaxies.  These have all been kinematically modelled using the WALLABY Kinematic Analysis Proto-Pipeline (\wkapp; \citealt{Deg2022}).

\textcolor{black}{Already, the WALLABY kinematic models are well suited to the study of scaling relations as they comprise the largest sample of \textit{untargetted, uniformly observed and analyzed} \hi\ disks available to date.  However, this sample still poses unique challenges for scaling relation studies as the majority of the models are for marginally to moderately resolved galaxies. Extracting \textcolor{black}{\hi\ disk structural properties} like the disk size with robust measures of the uncertainty from marginally resolved galaxies is not trivial. }

\textcolor{black}{This work develops the framework for extracting robust \textcolor{black}{\hi\ disk structural properties} and uncertainties from kinematic models of marginally resolved galaxies. It lays out a scalable approach that can be applied to the full WALLABY survey.  Using the combined WALLABY PDR1 and PDR2 set of kinematic models (hereafter PDR),} \textcolor{black}{we obtain measures of the \hi\ disk size, rotation velocity, and angular momentum. We combine them with  \hi\ mass measurements to construct the size-mass, size-velocity, \textcolor{black}{mass-velocity}, and angular momentum-mass relations. We also obtain stellar mass measurements allowing for the construction of the stellar TFR (sTFR), baryonic TFR (bTFR), the gas fraction-disk stability, and the gas fraction-baryonic mass relations.}

This paper is organized as follows: in Sec. \ref{sec:Data}, we present the details of the WALLABY observations and our methodology for calculating \textcolor{black}{the galaxy properties}.  \textcolor{black}{Then Sec. \ref{Sec:GasOnlyRelations} examines scaling relations that depend only on the \hi\ derived galaxy properties}.   Sec. \ref{Sec:bTFR} examines the sTFR and bTFR.  \textcolor{black}{And Sec. \ref{Sec:QAnalysis} examines the gas fraction-disk stability, and the gas fraction-baryonic mass relations.}  Finally Sec. \ref{Sec:Conclusions} presents our discussion and conclusions.

\section{Data}\label{sec:Data}

The currently released \textcolor{black}{WALLABY} PDR1 \citep{Westmeier2022} and \textcolor{black}{soon to be released PDR2} \citep{Murugeshan2024} \textcolor{black}{consist of fields} centered on the Hydra cluster, the Norma cluster, the NGC 4636 group, the Vela cluster, the NGC 4808 group, and the NGC 5044 group.   Kinematic modelling has been attempted for all marginally resolved galaxies ($\gtrapprox 3$ beams across) using the WALLABY Kinematic Analysis Proto-Pipeline (\wkapp; \citealt{Deg2022}).  Between PDR1 and PDR2, a total of 236 galaxies have been successfully modelled by \wkapp. \textcolor{black}{It is worth noting that the PDR fields are overdense with many galaxies lying in group and cluster environments, which means many of the galaxies may be disturbed by interactions, ram pressure stripping, and more \citep{Reynolds2021,Reynolds2022,Lee2022}.  These galaxies are generally not included \textcolor{black}{in the} final sample of 236 kinematic models for two reasons\textcolor{black}{:} \wkapp\ cannot fit strongly disturbed galaxies, and the visual inspection step removes those galaxies that are not be fully described as a rotating disk. }

As described in \citet{Deg2022}, \wkapp\ fits galaxies using the Fully Automated TiRiFiC code (\fat; \citealt{Kamphuis2015}, which itself is built around the Tilted Ring Fitting Code, \textsc{TiRiFiC}; \citealt{Jorza2007}) and the 3D-Based Analysis of Rotating Objects From Line Observations \citep[\barolo;][]{diTeodoro15}. For optimization, \wkapp\ first smooths the WALLABY observations to a $12~\kms$ resolution before fitting. The two codes are set to only fit `flat disk' models where the geometric parameters (such as position angle and \textcolor{black}{inclination}) are constant as a function of galactocentric radius. 
\textcolor{black}{In order for a \wkapp\ fit to be successful, first both \barolo\ and \fat\ must individually fit the galaxy.  Then the fits are visually inspected to determine whether the both fits are successful and consistent with each other.  If both are indeed determined to be successful, the two fits are averaged together to produce a final \wkapp\ model. }
The \wkapp\ geometric parameters are the average of the fit values from \fat\ and \barolo\ with the uncertainty set to half difference between the two fit parameters. 
The reason for using the difference between the codes is that it is typically larger than the errors reported by the individual codes \citep{Deg2022}.
The two rotation curves (RCs) are then corrected to the model inclination, and these inclination corrected RC's are averaged together to produce the output RC and uncertainties. 

The procedure for \hi\ surface density (SD) profile is slightly different as \textsc{FAT} and \textsc{3DBAROLO} fit the SD profiles in fundamentally different manners. In \wkapp, the output SD profile is constructed by ellipse fitting the WALLABY moment 0 map using the model geometry. As such, the \wkapp\ SD profile for a system can extend beyond the model RC.  These \textit{projected} SD profiles can be corrected to face-on profiles using a simple $\cos(i)$ approximation, \textcolor{black}{which we apply using the inclination returned by \wkapp.}  \textcolor{black}{While this inclination correction is} reliable \textcolor{black}{for WALLABY detections} in the outer SD profile when coupled with a beam correction, it fails in the beam-smeared inner regions \citep{Deg2022,Halloran2023}.

In what follows, the extraction of the \hi\ mass \textcolor{black}{$M_{\hi}$, \hi\ disk diameter $D_{\hi}$ and scale velocity $V_{\hi}$,} and their associated uncertainties from WALLABY pilot kinematic models is described.  \textcolor{black}{A machine readable table with all \textcolor{black}{galaxy properties} is available in the online Journal. The various columns, units, and descriptions for the data table are listed in Table \ref{tab:DataTableColumns}}.  
\textcolor{black}{The distances to the WALLABY galaxies are calculated in one of two ways.  Galaxies with systemic velocities less than $5000~\kms$ have distances derived from the Cosmicflows-3 Distance-Velocity Calculator \citep{Kourkchi2020}.  For galaxies with systemic velocities greater than $5000~\kms$ we use the Hubble flow distances (with $H_{0}=70~\kms\mathrm{Mpc^{-1}}$) calculated using the CMB corrected systemic velocities \citep{Fixsen1996}. }

\begin{table*}
    \centering
    \begin{tabular}{ c c c c }
        \hline
        Column   & Name                            & Units                      & Description \\
        \hline  
        1        &   WALLABY                &                            & The WALLABY designated name of the galaxy.\\
        2        &   RAdeg-model            & $^{\circ}$                 & The kinematically modelled center of the galaxy from \wkapp. \\
        3        &   DEdeg-model            & $^{\circ}$                 & The kinematically modelled center of the galaxy from \wkapp. \\
        4        &   Vsys-model             & $\kms$                     & The kinematically modelled systemic velocity of the galaxy from \wkapp. \\
        5        &   FCorr                  & Jy Hz                      & The corrected total flux of the galaxy. \\ 
        6        & f\_Dist                  &                            & A flag indicating how the distance is calculated. \\
        7        &   Dist                   & Mpc                        & The calculated distance to the galaxy. \\       
        2        &   I                      & $^{\circ}$                 & The measured galaxy inclination from \wkapp.\\
        3        & e\_I                     & $^{\circ}$                 & The uncertainty on the galaxy inclination from \wkapp.\\
        5        &   DHIa                   & \arcsec                    & $D_{\hi}$ of the galaxy in arcseconds.\\
        6        & b\_DHIa                  & \arcsec                    & The lower limit on $D_{\hi}$ in arcseconds.\\
        7        & B\_DHIa                  & \arcsec                    & The upper limit on $D_{\hi}$ in arcseconds.\\
        8        &   DHIk                   & kpc                        & $D_{\hi}$ of the galaxy in kiloparsecs.\\
        9        & b\_DHIk                  & kpc                        & The lower limit on $D_{\hi}$ in kiloparsecs.\\
        10       & B\_DHIk                  & kpc                        & The upper limit on $D_{\hi}$ in kiloparsecs.\\  
        11       &   VHI                    & $\kms$                     & $V_{\hi}$ of the galaxy .\\
        12       & e\_VHI                   & $\kms$                     & The lower uncertainty value for $V_{\hi}$.\\
        13       & E\_VHI                   & $\kms$                     & The upper uncertainty value for $V_{\hi}$.\\  
        14       &   Jcross                 & kpc $\kms$                 & $j_{\rm{X}\hi}$ of the galaxy.\\
        15       & e\_Jcross                & kpc $\kms$                 & The lower uncertainty value for $j_{\rm{X},\hi}$.\\
        16       & E\_Jcross                & kpc $\kms$                 & The upper uncertainty value for $j_{\rm{X},\hi}$.\\
        17       &   logMHI                 & $\log_{10}(M/\Msol)$       & The \hi\ mass of the galaxy corrected for the flux discrepancy \\
                 &                          &                            & and CMB distance.\\
        18       & e\_logMHI                & $\log_{10}(M/\Msol)$       & The lower uncertainty value for $M_{\hi}$.\\
        19       & E\_logMHI                & $\log_{10}(M/\Msol)$       & The upper uncertainty value for $M_{\hi}$.\\ 
        20       &   logMstar               & $\log_{10}(M/\Msol)$       & The stellar mass of the galaxy.\\
        21       & e\_logMstar              & $\log_{10}(M/\Msol)$       & The uncertainty in the stellar mass. \\
        22       &   logMBary               & $\log_{10}(M/\Msol)$       & The baryonic mass of the galaxy.\\
        23       & e\_logMBary              & $\log_{10}(M/\Msol)$       & The lower uncertainty value for the baryonic mass.\\
        24       & E\_logMBary              & $\log_{10}(M/\Msol)$       & The upper uncertainty value for the baryonic mass.\\  
        25       &   Photi                  & $^{\circ}$                 & The optical inclination of the galaxy. \\ 
        26       &   Qcross                 &                            & The disk stability parameter, $q_{\rm{X}}$.\\
        27       & E\_Qcross                &                            & The lower uncertainty value for $q_{\rm{X}}$.\\
        28       & e\_Qcross                &                            & The upper uncertainty value for $q_{\rm{X}}$.\\ 
        29       &   fatm                   &                            & The atomic gas fraction for the galaxy, $f_{\rm{atm}}$.\\
        30       & e\_fatm                  &                            & The lower uncertainty value for $f_{\rm{atm}}$.\\
        31       & E\_fatm                  &                            & The upper uncertainty value for $f_{\rm{atm}}$.\\ 
        32       &   RobustSampleFlag       &                            & A flag indicating if the galaxy has robustly measured \textcolor{black}{structural properties}.\\
        33       &   FullStellarFlag        &                            & A flag indicating if the galaxy belongs \\
                 &                          &                            & to the full stellar sample. \\
        34       &   InclMatchedFlag        &                            & A flag indicating if the galaxy belongs \\
                 &                          &                            & to the inclination matched stellar sample.\\     
        \hline 
    \end{tabular}
    \caption{The column names, units, and descriptions for the \textcolor{black}{galaxy properties} table. 
             The full machine readable table is available in the online Journal or downloadable from arXiv as an ancillary file.}
    \label{tab:DataTableColumns}
\end{table*}

\subsection{\hi\ Masses}\label{ssec:MassEtraction}

\textcolor{black}{We calculated the \hi\ mass of the WALLABY galaxies from the WALLABY PDR1 and PDR2 fluxes, which are themselves calculated during the source finding analysis \citep{Westmeier2022}.  These fluxes are generally lower than single dish observations of the same galaxies due to issues in the data cleaning and the dirty beam \citep{Westmeier2022}.  We apply the statistical flux corrections recommended by \citet{Westmeier2022} for PDR1 galaxies and \citet{Murugeshan2024} for PDR2 galaxies.
}

\textcolor{black}{The \hi\ masses of the kinematically modelled galaxies are}
\begin{equation}
    M_{\hi}/\Msol=49.7 F_{\rm{corr}} d^{2}/[\rm{Jy~Hz~Mpc}^{2}]~,
\end{equation}
where $F_{\rm{corr}}$ is the corrected flux in units of $\rm{Jy~Hz}$ and $d$ is the distance in Mpc.  
We then propagate the uncertainty in the uncorrected fluxes to uncertainties in the corrected \hi\ masses by keeping the relative uncertainty constant between the uncorrected flux and calculated \hi\ mass \textcolor{black}{(i.e. a 10\% uncertainty in the uncorrected flux measurement will lead to a 10\% uncertainty in the calculated \hi\ mass).  It is also worth noting that we do not propagate uncertainties in the distances into mass uncertainties due to the non-linear relationship between velocity and distance caused by the use of CF3 distances and CMB corrected distances.}

\subsection{$D_{\hi}$ Extraction} \label{ssec:RHICalc}

The \hi\ diameter of a galaxy is twice the \hi\ radius, $R_{\hi}$, which itself is defined as the point where the de-projected \hi\ SD profile reaches a threshold of $1~\rm{M}_{\odot}~\rm{pc}^{-2}$. 
This can be determined from the \wkapp\ SD profiles, but there are some complications to consider.
The \wkapp\ profiles are subject to beam smearing, so the point where they cross the $1~\Msol~\rm{pc^{-2}}$ limit is the uncorrected \hi\ radius, $R_{\hi,uc}$.  We determine $R_{\hi,uc}$ via linear interpolation between the outermost pair of points in the deprojected \wkapp\ SD profile bounding $1~\rm{M}_{\odot}~\rm{pc}^{-2}$.
Once the uncorrected radius is extracted, the corrected diameter can be found via the beam smearing correction of \citet{Wang2016}:
\begin{equation}\label{Eq:BeamCorr}
    D_{\hi}^{2}=D_{\hi,uc}^{2}-B^{2}~,
\end{equation}
 $B=30\arcsec$ is the full-width-at-half-maximum of the WALLABY clean beam.  

Estimating the uncertainty on $D_{\hi}$ is nontrivial due to uncertainties in both the inclination and \textit{projected} SD profile.  The procedure adopted for this work is:
\begin{enumerate}
    \item Calculate the upper and lower \textit{projected} SD profiles via the addition/subtraction of the \wkapp\ profile uncertainties.
    \item Deproject these profiles using the minimum and maximum inclinations via
    \begin{equation}
        \Sigma_{\rm{min/max}}=\Sigma_{\rm{min/max,projected}}\cos \left(i\pm \epsilon_{i} \right)~,
    \end{equation}
    where $i$ is the inclination and $\epsilon_{i}$ is \textcolor{black}{its} uncertainty.  Note that the combination of the minimum profile with the maximum inclination, and vice versa, ensures that the deprojected curves are truly the minimum and maximum possible.
    \item Extract the uncorrected upper and lower limits on $D_{\hi}$ using the maximum and minimum de-projected SD profiles.
    \item Correct these upper and lower limits for beam smearing using Eq. \ref{Eq:BeamCorr}.
\end{enumerate}

\begin{figure*}
\centering
    \includegraphics[width=0.9 \textwidth]{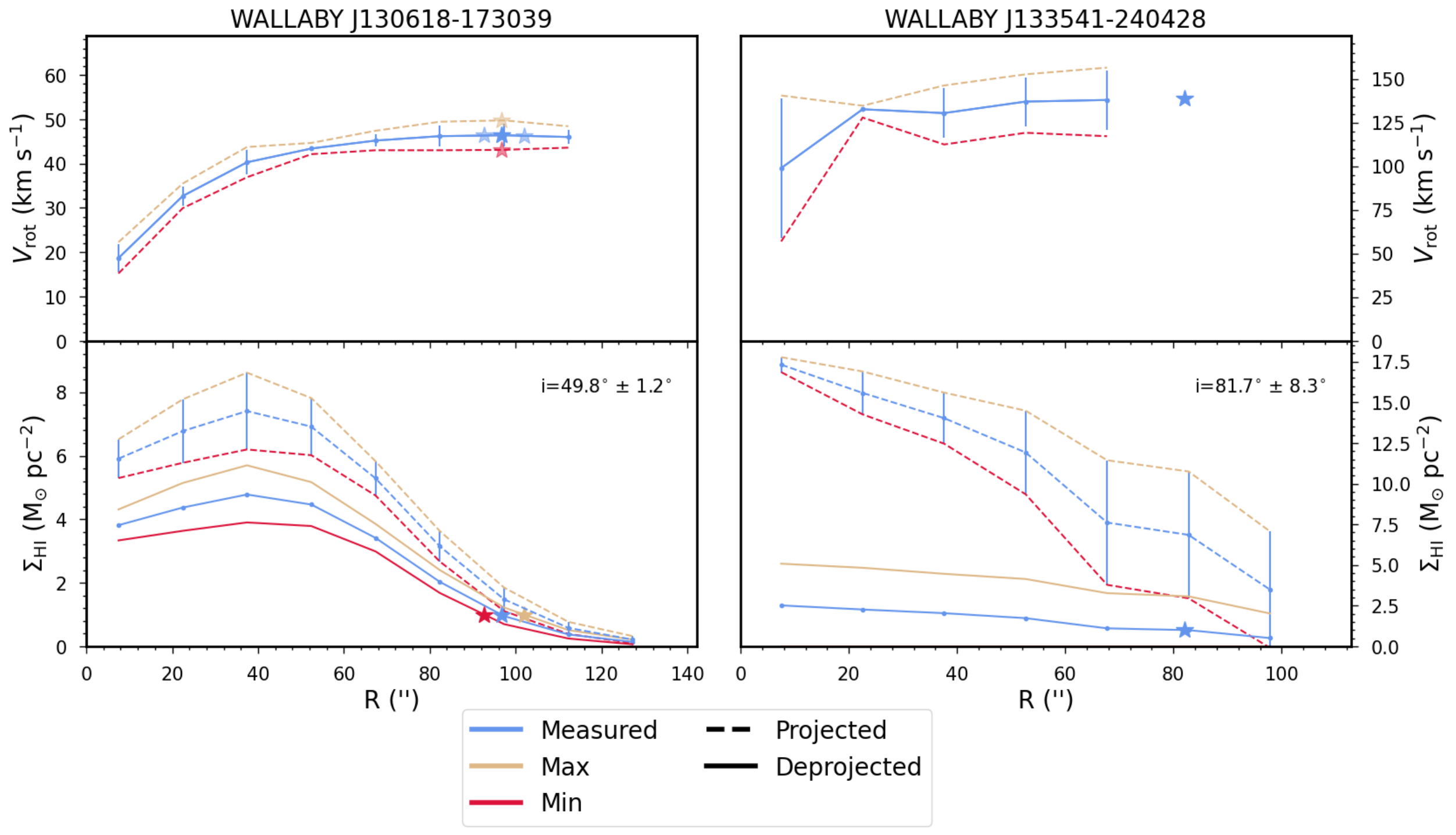} 
    \caption{\textcolor{black}{A pair of galaxies (WALLABY J130618-173039, left; WALLABY J133541-240428, right) demonstrating the extraction of the uncorrected $R_{\hi,uc}$ and $V_{\hi,uc}$ values using the model RCs (top) and SD profiles (bottom)}.   \textcolor{black}{For WALLABY J130618-173039 the parameter extraction is successful, while for WALLABY J133541-240428 the extraction fails and the galaxy is removed from the final samples.}
The blue, orange, and red lines show the measured, maximum, and minimum curves respectively.  
    In the lower panels, \textcolor{black}{the projected and deprojected profiles are shown as dashed and solid lines, respectively,} the blue star indicates $R_{\hi,uc}$ at which $\Sigma=1~\textrm{M}_{\odot}~\textrm{pc}^{-2}$ while the orange and red stars show the upper and lower limits on $R_{\hi,uc}$.  \textcolor{black}{In the lower right panel the minimum deprojected SD profile is $\approx 0~\Msol~\rm{pc^{-2}}$ at all radii due to the high inclination of the galaxy.  As such the solid red line cannot be seen due to the axes.}
In the upper panel, the dark blue star shows $V_{\hi}$, while the blue, orange, and red star triplets show the velocity at $R_{\hi}$ as well as the upper and lower limits on $R_{\hi,uc}$ for the measured, maximum, and minimum rotation curves respectively. \textcolor{black}{Note that the blue star showing $V_{\hi}$ for WALLABY J133541-240428 (right panel) is only intended to demonstrate that calculating $V_{\hi}$ can require extrapolation of the RC for some galaxies.}}
  \label{Fig:RHI_Sample}
\end{figure*}

\textcolor{black}{The bottom row of Figure \ref{Fig:RHI_Sample} illustrates the extraction of $R_{\hi,uc}$ and limits.} The galaxy shown in the lower-left panel, WALLABY J130618-173039, is well behaved and the extraction of $R_{\hi,uc}$ faces no issues.  But the galaxy shown in the lower right panel, WALLABY J133541-240428, is very different.   In that case, the maximum deprojected SD profile does not drop below $\Sigma=1~\textrm{M}_{\odot}~\textrm{pc}^{-2}$. Similarly, the minimum deprojected SD profile is never above $\Sigma=1~\textrm{M}_{\odot}~\textrm{pc}^{-2}$.  As such, we cannot determine reliable limits on the uncertainties for this source.  We discard all sources where either the lower or upper limit on $D_{\hi,uc}$ is not found.  \textcolor{black}{Finally the remaining sizes are converted to kpc using the CMB corrected distances discussed in Sec. \ref{ssec:MassEtraction}.}

\subsection{$V_{\hi}$ Extraction} \label{ssec:VHICalc}

\textcolor{black}{There are a variety of different rotational velocity definitions that can be used to characterize a galaxy. The most common is the velocity at the flat portion of the RC, $V_{\rm{flat}}$.  Others include $V_{2.2}$, $V_{2R_{e}}$, $V_{\rm{max}}$, and $V_{\rm{out}}$, which are the velocities at 2.2 times an exponential disk scale length, at 2 times the radius which encloses half the light, at the maximum of the RC, and at the last point in the outer RC respectively (see \citealt{Catinella2007,Lelli2019} for a more in depth discussion of these velocities).  Each of these have different motivations and strengths.  More importantly, the use of different methods of characterizing the galaxy velocity can change various scaling relations like the bTFR \citep{Lelli2019}.}

\textcolor{black}{In this work we use $V_{\hi}$, the velocity of the galaxy at $R_{\hi}$, to characterize a galaxy's rotation.  There are a number of advantages to this definition.  It does not presuppose a shape to the RC (e.g. \citealt{Catinella2008}) or SD profiles, it is tied explicitly to the disk size (allowing for a straightforward propagation of uncertainties), it is located at a specific isodensity contour meaning that it should probe the same portion of every galaxy, and it can be used for models with very few RC points (where the calculation of $V_{\rm{flat}}$ is difficult).  Once $D_{\hi}$ and its associated limits are calculated, $V_{\hi}$ and its uncertainty, $\epsilon_{V_{\hi}}$ can be calculated by linearly interpolating the RC to obtain the velocity at $R_{\hi,uc}$ (the use of $R_{\hi,uc}$ removes the need of beam correcting the entire RC before interpolating to get $V_{\hi}$).}

Determining the uncertainties in $V_{\hi}$ is slightly more complex as there are three distinct sources: the model RC uncertainty, the uncertainty in the RC due to the model inclination, and the uncertainty of $R_{\hi}$.  The first two sources of uncertainty \textcolor{black}{can be combined analogously to the approach for $R_{\hi,uc}$ in Sec.~\ref{ssec:RHICalc}:}
\begin{equation}
    V_{\rm{max/min}}(R)=\left[V(R)\pm \epsilon_{V}(R)\right]\frac{\sin(i\pm\epsilon_{i})}{\sin(i)}~.
\end{equation}
Then the maximum and minimum model velocities at $R_{\hi}$ can be calculated by linearly interpolating the maximum and minimum RCs at $R_{\hi}$.  The combined model uncertainty, $\epsilon_{V_{\hi},\rm{model}}$, is the average of the absolute differences of these maximum and minimum model velocities from $V_{\hi}$.

The third source of uncertainty to consider is that of $R_{\hi,uc}$.  To account for this, we calculate the velocities of the model at the lower and upper limits of $R_{\hi,uc}$.   Then the uncertainty on $V_{\hi}$ due to the uncertainty in $R_{\hi}$, $\epsilon_{V_{\hi},R_{\hi}}$, is set as the average of the absolute differences of these velocities from $V_{\hi}$.  With both terms calculated, the total uncertainty for $V_{\hi}$ is
\begin{equation}
    \epsilon_{V_{\hi}}^{2}=\epsilon_{V_{\hi},R_{\hi}}^{2}+\epsilon_{V_{\hi},\rm{model}}^{2}~.
\end{equation}

While the method of calculating $V_{\hi}$ uncertainties is robust, \textcolor{black}{they may be artificially low because the \wkapp\ RC uncertainties arise from the difference between the \fat\ and \barolo\ fits (see Sec. \ref{sec:Data}): when the fits are similar, their relative difference will be low, leading to very low uncertainties in $V_{\hi}$.  Those $V_{\hi}$ measures with very low uncertainties then pull strongly on the fits presented in Sec.~\ref{Sec:ScalingRelations}.}  In order to account for low $V_{\hi}$ uncertainties, we have added $6~\kms$ in quadrature to $\epsilon_{V_{\hi}}$ for all galaxies.  The $6~\kms$ term is half the size of the smoothed $12~\kms$ channels used in the \wkapp\ modelling.

The upper panels of Fig. \ref{Fig:RHI_Sample} show this extraction for WALLABY J130618-173039 and WALLABY J133541-240428. In the case of WALLABY J130618-173039 (upper left panel), the extraction goes well and $V_{\hi}$ is well constrained.  But, because the limits on $R_{\hi}$ are unknown for WALLABY J133541-240428, the uncertainty in $V_{\hi}$ cannot be calculated.  

\textcolor{black}{In \wkapp\, it is possible for the SD profile to extend further than the RC for a given galaxy}
Such is the case for WALLABY J133541-240428 (the right-hand panels of Figure \ref{Fig:RHI_Sample}).  Calculating $V_{\hi}$ for these cases requires extrapolating the RC, which is not necessarily accurate. Since the accuracy of the extrapolation decreases \textcolor{black}{with its extent}, we restrict our sample to those galaxies where $R_{\hi}-R_{\rm{outer,RC}} < 0.5$ beams, where $R_{\rm{outer,RC}}$ is the outermost radial point of the RC.  Because we wish to preserve gradient information, linear extrapolation of the last two points in the RC to determine $V_{\hi}$ (as well as the velocity at the upper limit of $R_{\hi}$) is used.  \textcolor{black}{Ultimately, the extrapolation} limit balances the need for statistically meaningful samples with the accuracy of the extrapolation.  The \textcolor{black}{resulting} sample \textcolor{black}{consists of 148 galaxies that satisfy the requirements of well measured sizes and velocities}. 21 of these 148 galaxies require \textcolor{black}{minimal extrapolation of their RCs to obtain $V_{\hi}$}. \textcolor{black}{It is important to note here that we have performed the rest of the analysis excluding those 21 galaxies, but found no difference to the recovered scaling relations.  This lack of a change indicates that the minimally extrapolated $V_{\hi}$ values are reliable, and including them improves the statistics of the analysis.}

\textcolor{black}{As a consistency check, it is worth comparing our $V_{\hi}$ values with the corrected \hi\ profile widths presented in \citet{Courtois2023} in their study of the TFR for WALLABY PDR1. For their study, they use \textcolor{black}{an inclination, redshift, turbulence, and instrumental broadening} corrected profile width, $W_{mx}$ (see \citet{Courtois2023} for details).  It is important to note that for this comparison there is not a perfect overlap between PDR1 galaxies with measured $V_{\hi}$ and $W_{mx}$ values as not all galaxies are kinematically modelled and not all of those that are modelled have robustly measured $V_{\hi}$ values, nor are all spectra of sufficient quality for $W_{mx}$ extraction.}

\begin{figure}
\centering
    \includegraphics[width=0.95\columnwidth]{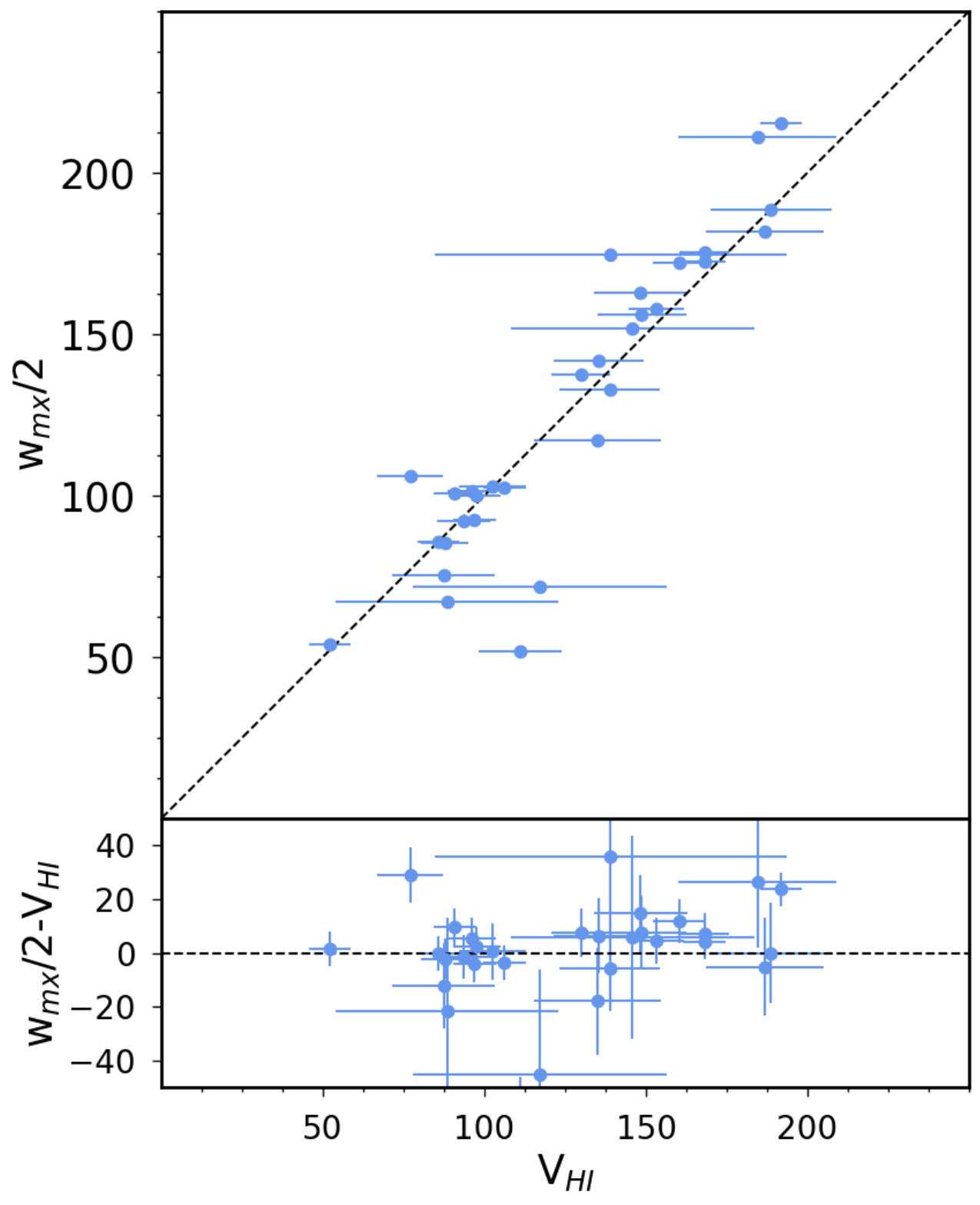} 
    \caption{A comparison of $V_{\hi}$ (x-axis) to the profile width, $W_{mx}$ (y-axis), of \citet{Courtois2023}. \textcolor{black}{The top panel shows both values with the dashed black line indicating the 1-1 line. The bottom panel shows the difference between the two with $V_{\hi}$ uncertainties, where the dashed black line indicates a difference of zero}.}
  \label{Fig:VelComp}
\end{figure}

\textcolor{black}{Figure \ref{Fig:VelComp} shows the correlation of their widths $W_{mx}/2$ to $V_{\hi}$ for the 31 PDR1 galaxies with both robust $V_{\hi}$ and $W_{mx}$ measurements. The two velocity measurements broadly agree, although the $W_{mx}$ velocities are slightly higher in the mean than $V_{\hi}$, likely due to spectral broadening.  The few outliers are most likely caused by differences in the inclinations used in the kinematic models and the profile analysis.  Given this close correspondence, we are confident that our $V_{\hi}$ is appropriate for studies of scaling relations.}

\subsection{Angular Momentum}\label{ssec:AngMomCalc}

\textcolor{black}{Studies of well-resolved \hi\ disks \citep[e.g.][]{Kurapati2021,ManceraPina2021a,Sorgho2024} typically measure specific angular momentum $j$ as}
\begin{equation}\label{Eq:specAM}
    j=\frac{\int_{0}^{R}2\pi R^{'2}\Sigma(R^{'}) V(R^{'}) dR}{\int_{0}^{R}2\pi R^{'} \Sigma(R^{'}) dR^{'}}~,
\end{equation}
\textcolor{black}{where $\Sigma(R^{'})$ is the SD profile and $V(R^{'})$ is the RC. However, in the marginally resolved regime, the calculation of $\Sigma(R^{'})$ in the innermost regions is uncertain due to beam smearing (even 3D modelling is susceptible to beam smearing in the innermost region).  And, when the surface density profile is derived from ellipse fitting, the $\cos(i)$ correction is also unreliable in the inner region \citep{Deg2022,Halloran2023}.  Instead, we use \textcolor{black}{a} simpler $j_{\rm{X},\hi}$ defined as}
\begin{equation}
\label{Eq:am}
    j_{\rm{X},\hi}=\frac{1}{2}D_{\hi}  V_{\hi}~.
\end{equation}
\textcolor{black}{This is effectively an upper limit on the specific angular momentum, which treats the system as though all the mass is located at $R_{\hi}$.  As it is defined by the scale radius and velocity, the uncertainties can be calculated by propagating the relative error in the relevant quantities to $j_{\rm{X},\hi}$.}

\subsection{Stellar Masses}\label{ssec:StellarMass}
\NA{An important addition to the gas properties of galaxies is information about the distribution of stars retrieved through multi-band photometry.}
\textcolor{black}{In this work, we use} the deep optical imaging of the Dark Energy Sky Instrument Legacy Imaging Survey Date Release-10 \citep[hereafter DESI;][]{DESI, Dey2019} within the \textit{grz} bands to extract stellar masses for the WALLABY \textcolor{black}{galaxies in the DESI footprint}.  

For each galaxy, the surface brightness (SB) profile in each of the DESI \textit{grz} photometric bands is calculated using the \textsc{AutoProf} package \citep{Stone2021}.  \textsc{AutoProf} examines an image and performs a non-parametric azimuthally-averaged decomposition to obtain radial profiles of position angles (PA), ellipticity and SB.
\NA{Owing to its high signal-to-noise ratio and minimal dust extinction, the optical centres, PA, and ellipticity profiles are calculated using the DESI-\textit{r} band image. To obtain SB profiles in \textit{gz}-bands, we apply forced photometry using the DESI-\textit{r} geometric profiles to ensure a uniform extraction in all bands.}

Our procedure for optical stellar mass estimates \textcolor{black}{is similar to that of} \cite{Arora2021}.
Firstly, \textit{grz} SB profiles are truncated once an SB \textcolor{black}{uncertainty} of 0.22 mag\,arcsec$^{-2}$ is \textcolor{black}{reached}. Then the light inside that region is used to calculate the global luminosities and optical colors are calculated.
The stellar mass-to-light ratios ($\rm \Upsilon_*$) for converting the luminosity to the stellar mass estimates are calculated using various mass-to-light color relations \citep[MLCRs;][]{Courteau2014}.
Rather than use a single MLCR, we use five different MLCRs from \cite{Roediger2015}, \cite{Zhang2017}, and \cite{Benito2019} which differ in assumptions about stellar population synthesis model, star formation history, dust extinction, and calibration input data (global vs. resolved \textcolor{black}{spectral energy distributions}). Each of these MLCRs can be used to obtain a $\rm \Upsilon_*$ using any of the three colors ($g-r$, $g-z$, and $r-z$).
The use of the multiple colors provides better constraints on $\rm \Upsilon_*$ estimates, reducing the random error on the final stellar mass measurement due to various assumptions in the different MLCRs. Each of the $\rm \Upsilon_*$ are multiplied by the luminosities from the  \textit{g}, \textit{r} and \textit{z} bands yielding 45 stellar mass estimates (5 MLCRs $\times$ 3 colours $\times$ 3 luminosities). 
\textcolor{black}{The adopted stellar mass is the median of these 45 measurements and its uncertainty is their standard deviation}.
\NA{In principle, mid-infrared (MIR) imaging can also be used to estimate stellar mass estimates; but in practice the results are independent of the flux tracer implemented \citep[see][for a comparison for optical-MIR $M_*$ comparison]{Arora2021}.
Therefore, our choice of using the optical photometry emerges from the high resolution and the surface brightness depth of the DESI imaging. 
}

While our procedure to extract optical stellar masses reduces the random error, it should be noted that a systematic error of $\sim 0.3\,{\rm dex}$ (related to assumptions about star formation history, dust extinction, calibration input data, etc.) still remains \citep{Courteau2014}. It is \textcolor{black}{also} important to note here that we are unable to obtain stellar masses for every kinematically modelled galaxy in this study.  \textcolor{black}{Some galaxies lie in the zone of avoidance and have no known optical counterpart, while for some other galaxies the DESI imaging is unavailable in all three bands.}   Ultimately, there are 92 galaxies with robust \hi\ \textcolor{black}{structural properties} and stellar masses.

\textcolor{black}{The stellar masses can be combined with the \hi\ masses to obtain baryonic masses for galaxies.  We set the baryonic mass, $M_{b}$ to}
\begin{equation}
    M_{b}=M_{*}+1.35M_{\hi}~,
\end{equation}
where $M_{\hi}$ is the \hi\ mass, $M_{*}$ is the stellar mass.  The factor of \textcolor{black}{1.35} is used to convert the \hi\ mass to a total gas mass including helium and metals \citep{Arnett1999,Oh2015}. 

\subsection{Gas Fraction and Stability}\label{ssec:QExtraction}

\textcolor{black}{The combination of stellar masses and angular momentum allows for probes of the gas fraction and disk stability that were posited in \citet{Obreschkow2016}.  The atomic gas fraction is }
\begin{equation}
    f_{\rm{atm}}=\frac{1.35 M_{\hi}}{M_{b}}~,
\end{equation}
\textcolor{black}{which means that it goes to 1 as the baryonic mass becomes gas dominated.  }

\textcolor{black}{
The `global disk stability' parameter $q$ introduced in \citet{Obreschkow2014}, which is related to the Toomre Q parameter \citep{Toomre1964}.  Explicitly this parameter is defined as }
\begin{equation}
    q=\frac{j_{b} V_{\rm{disp}}}{GM_{b}}~,
\end{equation}
\textcolor{black}{
where $j_{b}$ is the baryonic specific angular momentum and $V_{\rm{disp}}$ is the gas velocity dispersion.  The baryonic specific angular momentum is obtained by integrating Eq. \ref{Eq:specAM} with the baryonic surface density profile rather than the gas surface density.  As discussed in Sec. \ref{ssec:AngMomCalc}, such integrations are unreliable for the majority of the WALLABY models, so we adopt an alternate stability parameter definition,
\begin{equation}
    q_{\rm{X}}=\frac{j_{\rm{X},\hi} V_{\rm{disp}}}{GM_{b}}~.
\end{equation}  
For the WALLABY PDR models, all velocity dispersions are fixed to $8~\kms$ \citep{Deg2022}. \textcolor{black}{This value was chosen based on the choice to Hanning smooth the cubes to $12~\kms$ prior kinematic modelling as well as the results of \citet{Tamburro2009} who found that galaxies typically have a tight range of gaseous velocity dispersions of $8~\kms\le V_{\rm{disp}}\le 12~\kms$ (see also \citealt{Iorio2017,Bacchini2019,ManceraPina2021a}). Since $j_{\rm{X},\hi}$ is an upper limit on the gaseous angular momentum, $q_{\rm{X}}$ is the maximal or outer stability of a pure gas disk.  } }

\subsection{Samples and Properties}\label{ssec:Selection}

\begin{figure}
\centering
    \includegraphics[width=0.95\columnwidth]{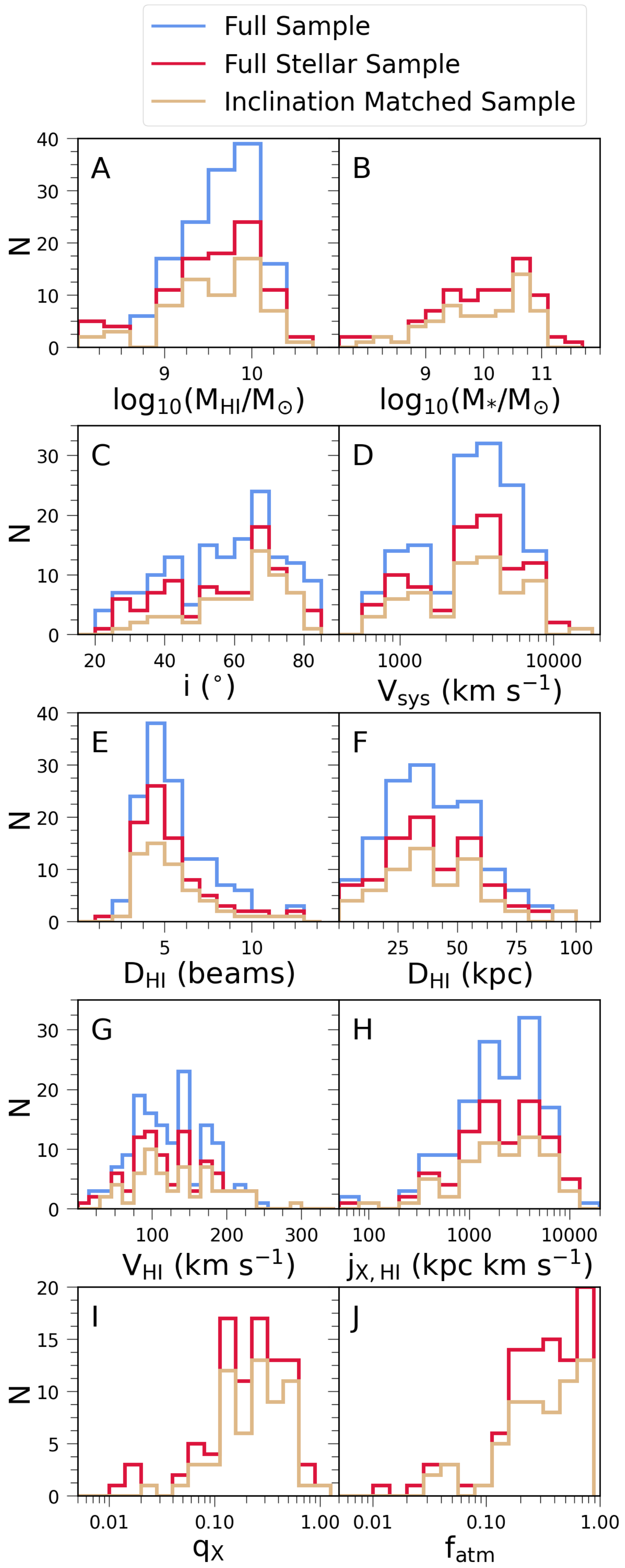} 
    \caption{The distribution of the galaxy properties for the full sample (blue lines), full stellar sample (red lines), and the inclination matched stellar mass selected sample (gold lines). Note that panels B, I, and J do not include the full sample line as those properties require stellar masses.}  
\label{Fig:SampleHists}
\end{figure}

WALLABY PDR1 and PDR2 contains a total of 236 kinematic models. However, as seen in Figure \ref{Fig:RHI_Sample} and discussed in Sec. \ref{ssec:VHICalc}, not all the models extend out far enough for reliable calculations of the various scale parameters.  Restricting the sample to those with reliable $R_{\hi}$ uncertainties and only minimal velocity extrapolation for $V_{\hi}$ provides a sample of 148 galaxies. The distribution of the \textcolor{black}{galaxy properties} for these \textcolor{black}{systems} are presented \textcolor{black}{in Figure \ref{Fig:SampleHists} as blue lines.  Further, the set of 92 galaxies with both stellar masses and robust \textcolor{black}{structural properties} are shown as red lines.} 

The kinematically modelled population of WALLABY galaxies tend to be relatively massive, with the peak of the \hi\ mass distribution lying near $\sim 10^{9.8}~\Msol$ (panel A).  \textcolor{black}{Similarly, the stellar masses peak near $\sim 10^{10.5}~\Msol$ (panel B).  Moving to the inclination, there are systematics in \wkapp\ that lead to a bias towards higher inclination models (panel C) \textcolor{black}{due to biases in \fat\ and \barolo\ for marginally resolved galaxies with a single resolution element across the minor axis.} The distribution of systemic velocities is more interesting, with some of the large scale structure seen in the different fields appearing in panel E, implying that there is a bias towards galaxies living in cluster/group environments.  This is not surprising given that the PDR fields were chosen to point towards denser regions of space.  Also unsurprising given the untargetted nature of WALLABY, the majority of models are marginally resolved, with a peak around $D_{\hi}\sim 5$ beams (panel E).  But, when converted to kpc in panel F, the distribution broadens significantly.  Both the galaxy sizes and scale velocities (panel G) have slightly skewed distributions with the peaks shifted to lower values in linear space.  By contrast, the maximal specific angular momentum (panel H) seems skewed towards higher values.  This difference is a consequence of the angular momentum being plotted in logarithmic bins while the size and velocity parameters are in linear bins.  Finally, both the disk stability parameter (panel I) and atomic gas fraction (panel J) lie predominantly in the range $0.1-1.$
}

\textcolor{black}{The goal of the preceding work is to robustly measure \textcolor{black}{\hi\ disk structural properties} from \wkapp\ kinematic models.  However, there is a potential for the models themselves to have issues.  \wkapp\ is designed to be applied to WALLABY detections without considering optical and other ancillary data.  As a result, the kinematically inclination is based purely on the \hi, and it is subject to errors.  }

\textcolor{black}{The 92 galaxies with stellar mass measurements also have photometric inclinations derived from \textsc{AutoProf}.  These can be used to further constrain the sample to those galaxies where the two inclination measures agree.  For studies of the sTFR and bTFR one of the two key parameters is $V_{\hi}$, which is sensitive to $\sin(i)$.  Given this, as well as the fact that the original \wkapp\ modelling algorithm adopts  a $\Delta \sin(i)$ cut, we construct an inclination matched stellar sample using the additional constraint that $\Delta  \sin(i_{\rm{\wkapp}})-\sin(i_{\rm{optical}}) <0.1$.  This is a relatively conservative cut, but it ensures that issues with the inclination measurement are unlikely to be the cause of outliers in the relationships probed by this sample.  Ultimately, the inclination matched stellar sample consists of 61 galaxies, shown as gold lines in Figure \ref{Fig:SampleHists}.  These follow the same distribution as the full sample of 148 galaxies and full stellar sample of 92 galaxies.  As such, we do not expect the use of the inclination matched stellar sample to lead to biases in our results.  A brief summary of all three samples are listed in Table \ref{tab:Sample}.
}

\begin{table*}
    \centering
    \begin{tabular}{ c c c }
        \hline
        Sample Name       & Selection Criteria     & Number of galaxies \\
        \hline 
         Full Sample     & Robust measures of $D_{\hi}$ and $V_{\hi}$.       & 148 \\ 
         Full Stellar Sample       & Those galaxies in the Full Sample with stellar masses.                                                 & 92\\
         Inclination Matched Stellar Sample       & Those galaxies in the full stellar sample where                                                 & 61 \\
       &  the optical and kinematic inclinations agree.       & \\
         \hline
    \end{tabular}
    \caption{The various samples used for analysis of scaling relations.}
    \label{tab:Sample}
\end{table*}

\section{Gas Only Scaling Relations}\label{Sec:GasOnlyRelations}

Equipped with the mass, size, velocity, and angular momentum measurements and associated uncertainties for the population of kinematically modelled galaxies, the comparisons and examination of various \hi\ scaling relations can now be presented.  \textcolor{black}{In general, scaling relations are written as} 
\begin{equation}
    Y=\alpha X + \beta~,
\end{equation}
\textcolor{black}{where $X$ and $Y$ are the \textcolor{black}{galaxy properties} and $\alpha$ and $\beta$ are the slope and intercept respectively.  Typically the \textcolor{black}{galaxy properties} are logarithmic quantities, meaning that scaling relations are power laws. In this section we will examine these \hi\ scaling relations in both a qualitative (Sec. \ref{Sec:ExistingScalingRelations}) and quantitative (Sec. \ref{Sec:ScalingRelations}) fashion.} 

\subsection{Qualitative Comparison of \hi\ Scaling Relations}\label{Sec:ExistingScalingRelations}

\begin{figure*}
\centering
    \includegraphics[width=0.95\textwidth]{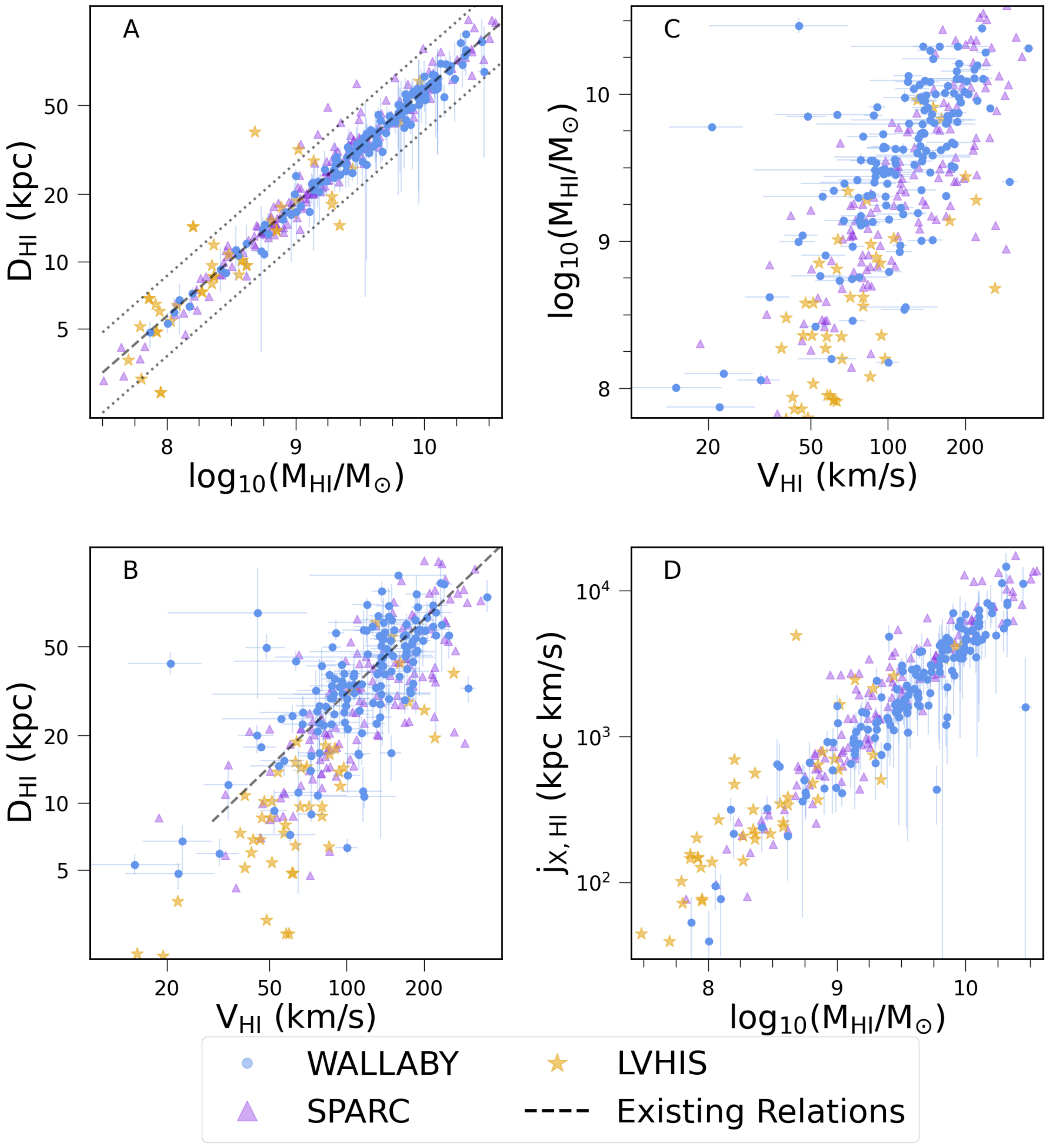} 
    \caption{The size-mass (panel A), size-velocity (panel B), \textcolor{black}{mass-velocity} (panel C), and angular momentum-mass (panel D) relations for the full sample of 148 galaxies from the WALLABY survey (blue circles with errorbars).  The purple triangles present equivalent measures from SPARC survey \citep{Lelli2016}, while the \textcolor{black}{gold} stars show those of LVHIS \citep{Koribalski2018}.  The dashed \textcolor{black}{and dotted} grey lines in panel A shows the \citet{Wang2016} size-mass relation \textcolor{black}{and associated $\pm3~\sigma$ limits}, while the dashed grey line in panel B shows the \citet{Meurer2018} size-velocity relation \textcolor{black}{over the range explored in that study.} }
\label{Fig:ScalingRelations_Full}
\end{figure*}

\textcolor{black}{Figure \ref{Fig:ScalingRelations_Full} shows four \hi\ scaling relations traced by the full sample of 148 WALLABY galaxies; the size-mass relation, the size-velocity relation, the \textcolor{black}{mass-velocity} relation, and the angular momentum-mass relation.  It compares these to existing relations (dashed grey lines) where applicable as well as to measures from other surveys.}

\textcolor{black}{The size-mass relation, which relates the diameter of the \hi\ disk to the total \hi\ mass, is one of the tightest scaling relations in all extragalactic astrophysics \citep{Broeils1997,Verheijen2001,Swaters2002,Noordermeer2005,Begum2008,Wang2013,Stevens2019}.}
\textcolor{black}{\citet{Wang2016} examined 562 galaxies from a variety of different surveys spanning \hi\ masses} \textcolor{black}{from $10^{5}~\Msol \lesssim \hi\ \lesssim 10^{11}~\Msol$, finding 
\begin{multline}\label{Eq:SizeMass_Wang}
    \log_{10}(D_{\hi}/{\rm kpc})=(0.506\pm 0.003)\log_{10}(\textrm{M}_{\hi}/\textrm{M}_{\odot})\\-(3.293\pm 0.009)~,
\end{multline}    
As shown in Panel A of Figure \ref{Fig:ScalingRelations_Full}, the 148 WALLABY kinematically modelled galaxies with robust scaling parameter measurements fall on the \citet{Wang2016} relation.}

\textcolor{black}{The non-WALLABY galaxies shown} in Figure \ref{Fig:ScalingRelations_Full} are from the SPARC survey \citep[purple triangles; ][]{Lelli2016} and the Local Volume \hi\ Survey (LVHIS; gold stars; \citealt{Koribalski2018}). \textcolor{black}{These surveys are complementary to WALLABY in terms of probing scaling relations.}  The SPARC sample consists of 175 nearby galaxies with robustly measured rotation curves and mid-infrared SD profiles.  \textcolor{black}{In \citet{Lelli2016}, the \hi\ masses, $R_{\hi}$ and $V_{\rm{flat}}$ (the flat portion of the RC) are reported for many of the galaxies.
The LVHIS sample consists of 82 nearby gas-rich galaxies in the local volume.  There are 47 galaxies for which \hi\ masses, $R_{\hi}$ and an outer rotation velocity, $V_{\rm{rot}}$ are available.  
}

The WALLABY galaxies, SPARC galaxies and LVHIS galaxies all show the same size-mass relation in Panel A of Figure \ref{Fig:ScalingRelations_Full}.  It is possible that the WALLABY galaxies have a smaller scatter than the LVHIS and SPARC galaxies likely due to larger uncertainties and measurement  nuances. \textcolor{black}{The LVHIS and SPARC samples reach lower \hi\ masses than the WALLABY sample;} \textcolor{black}{this is expected because the untargetted WALLABY pilot survey fields cover a limited sky area, which implies that the nearby volume sampled in which low-mass galaxies are detectable is very small \citep[e.g.][]{Giovanelli2015}.}

\textcolor{black}{The size-velocity relation has been extensively studied in the optical \citep{Courteau2007,Dutton2007, Avila-Reese2008, Saintonge2011, Ouellette2017, Lapi2018, Stone2021, Arora2023}.\citet{Meurer2018} instead examined the relation in \hi.  Among the parameterizations they explored, the most similar to our WALLABY sample is built using the outermost radius $R_{\rm{max}}$ and velocity $V(R_{\rm{max}})$ for modelled galaxies in the \citet{Meurer2013} sample:
\begin{multline}
    \log_{10}(R_{\rm{max}}/{\rm kpc})=(1.99\pm 0.24)\log_{10}(\textrm{V}_{R_{\rm{max}}}/\kms)\\-(1.10\pm 0.11)~.
\end{multline}
and shown as the dashed black line in Panel B of Figure \ref{Fig:ScalingRelations_Full}. 
}

A good agreement exists between the WALLABY data and the \citet{Meurer2018} relation. This is perhaps somewhat surprising as our $V_{\hi}$ and $R_{\hi}$ measures are not the same as the $R_{\rm{max}}$ and $V(R_{\rm{max}})$ estimates used by \citet{Meurer2018}.  \textcolor{black}{$R_{\rm{max}}$ is the outermost radial point of a RC and not an isodensity contour.  In cases where the RC extends far enough, $R_{\rm{max}}>R_{\hi}$, but, depending on the survey sensitivity, many \hi\ derived RCs may not extend out to $R_{\hi}$ (see for example the upper right hand panel of Figure \ref{Fig:RHI_Sample}).  Given the differences between these velocity definitions, and that they depend on the RC shape, and thus the galaxy mass \citep{Catinella2007,Catinella2008}, it is not expected \textit{a priori} that the two would agree .  The fact that both are giving similar relations suggests that $V_{\hi}$ and $V(R_{\rm{max}})$ of \citet{Meurer2018}} are sampling the flat portion of the RC.  Like the size-mass panel, there are certainly outliers, but these often have large uncertainties. 
However, the size-velocity relation has a significantly larger scatter than the size-mass relation.  
The WALLABY sample is broadly consistent with the SPARC sample as well.  \textcolor{black}{This agreement is noteworthy as their $V_{\rm{{flat}}}$ is not the same as our $V_{\hi}$.  It suggests that $V_{\hi}$ is typically located in the flat part of the RC, which was not necessarily true.  In point of fact, the LVHIS galaxies use outermost point in their rotation curves, $V_{\rm{rot}}$, to characterize their RCs and \textcolor{black}{at lower values of $V_{\rm{rot}}$}, those galaxies tend to lie below the WALLABY and SPARC galaxies as well as the \citet{Meurer2018} line.  \textcolor{black}{The lower velocity, lower mass LVHIS galaxies are more likely to have rising rotation curves \citep{Catinella2006}.  These curves, coupled with the use of $V_{out}$ rather than $V_{\hi}$, would tend to cause the LVHIS galaxies to lie below the SPARC and WALLABY samples.}}

\textcolor{black}{Next, we investigate the velocity-\hi\ mass relation, which is shown in Panel C of Figure \ref{Fig:ScalingRelations_Full}.  This relation, which is a gaseous counterpart to the sTFR and bTFR, has not received the same attention as those other \textcolor{black}{mass-velocity} relations.  Nonetheless, \citet{Brooks2017} and \citet{Chauhan2019} have studied it in simulations and observations and find a clear power-law, although \citet{Brooks2017} noted that mock gas-rich observations appear to be separated from gas poor observations.  In the case of WALLABY, the modelled galaxies are generally gas rich.  Neither of these provide an explicit relation, so no such relation is shown in Panel C.}
Broadly speaking, the WALLABY galaxies are again consistent with the SPARC and LVHIS samples \textcolor{black}{in this parameter space,} \textcolor{black}{although the latter suggest an upturn at lower \hi\ masses than probed by our WALLABY sample.}

\textcolor{black}{Panel D of Figure \ref{Fig:ScalingRelations_Full} presents the maximal specific angular momentum-\hi\ mass relation.  The specific angular momentum has been studied extensively \citep{Fall1980,Romanowsky2012,Obreschkow2014,Murugeshan2020,Kurapati2021,ManceraPina2021a,Elson2023} and is correlated with many physical parameters, including morphology,} \textcolor{black}{gas fraction, and disk stability. }

\textcolor{black}{
Our use of $j_{\rm{x},\hi}$ in the marginally resolved regime rather than the integrated specific angular momentum complicates direct comparisons with measures from previous studies. In order to compare WALLABY galaxies with SPARC and LVHIS galaxies, we calculate $j_{\rm{x},\hi}$ for each of those samples using their characteristic scale velocities ($V_{\rm{flat}}$ and $V_{\rm{rot}}$ for SPARC and LVHIS respectively).  Given that the three surveys agree for the other scaling relations, it is unsurprising that they also agree for the maximal specific angular momentum-mass relation.  More broadly, this relation is fairly tight over a large range of magnitudes, likely due to the tightness of the size-mass relation projected into angular momentum space.
}

\begin{figure}
\centering
    \includegraphics[width=\columnwidth]{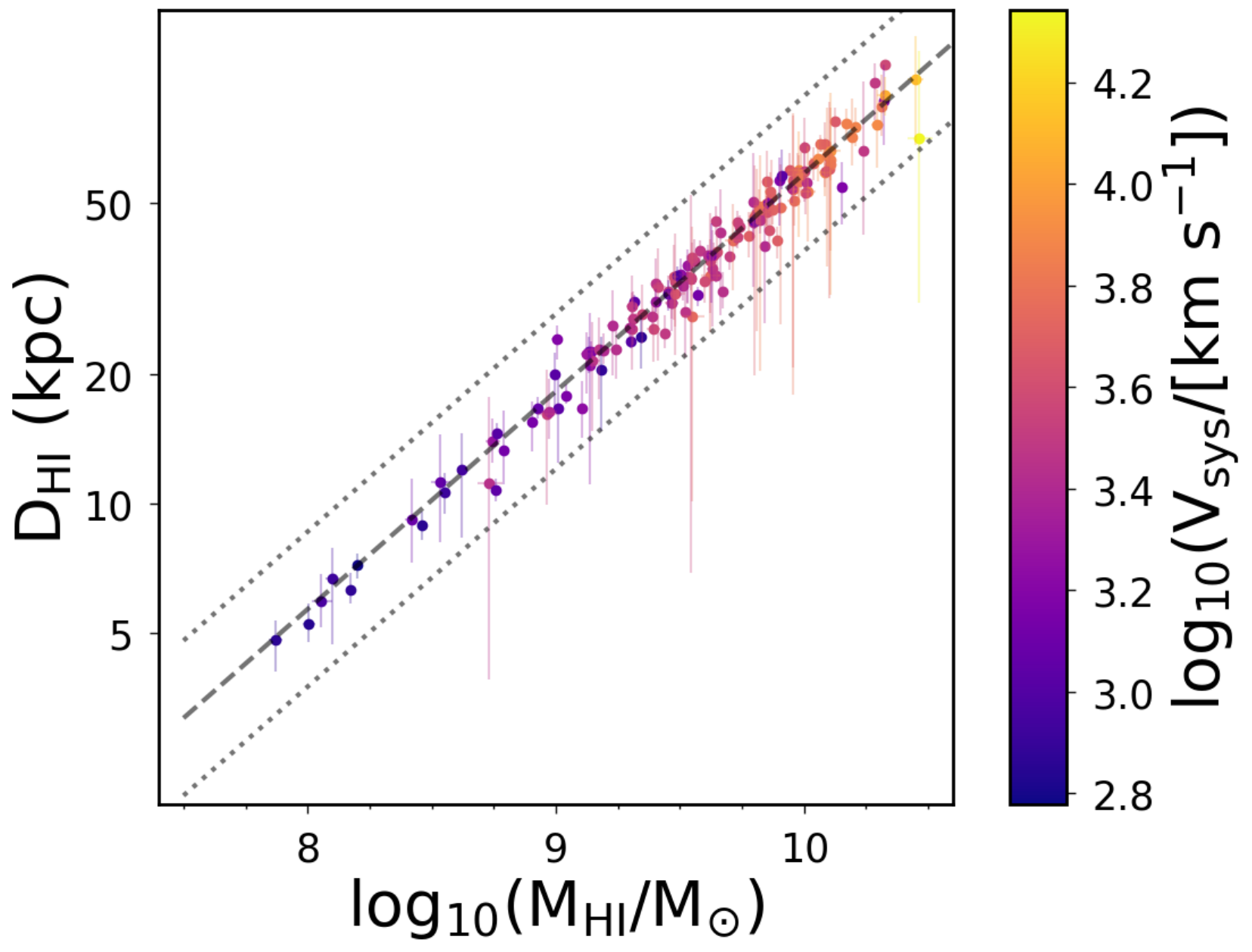} 
    \caption{The size-mass relation of WALLABY sample colored according to the modelled systemic velocity. The dashed and dotted lines grey lines are the \citet{Wang2016} relation and $\pm3~\sigma$ scatter respectively.}
  \label{Fig:ScalingRelations_Limited}
\end{figure}

\textcolor{black}{In order to investigate possible correlations and biases in the WALLABY sample, we re-plotted the size-mass relation and colored the galaxies by their various properties.  Figure \ref{Fig:ScalingRelations_Limited} shows the clearest correlation, which is the relation of size and mass to systemic velocity.}  This trend \textcolor{black}{arises from the combination of WALLABY's untargetted survey approach, the tight correlation of \hi\ mass and size (Figure~\ref{Fig:ScalingRelations_Full}A), and the resolution requirements of for kinematic modelling.}   Low \hi\ mass galaxies are intrinsically smaller, which means that only those in a smaller volume (with lower $V_{\rm{sys}}$ values) will have the observed resolution required for kinematic modelling, while higher mass galaxies will be remain resolved at greater distances. \textcolor{black}{Nearby high mass galaxies will be detected and modelled in WALLABY; however, since the volume sampled increases with distance, most high-mass galaxies are at higher $V_{\rm{sys}}$. This underscores the importance of quantifying the selection function in interpreting scaling relation results.}

\subsection{Quantitative Scaling Relations}\label{Sec:ScalingRelations}

\textcolor{black}{With qualitative trends in the \hi\ scaling relations for the WALLABY sample and existing scaling relations explored in Sec.~\ref{Sec:ExistingScalingRelations}, we now proceed to a quantitative analysis.  An important factor to consider are the outliers and the distribution of uncertainties.  As seen in Figures \ref{Fig:ScalingRelations_Full} and \ref{Fig:ScalingRelations_Limited}, many of these galaxies have large uncertainties.  However, these uncertainties have little constraining power and do not strongly affect the fits.  Of more concern are outlier galaxies with low uncertainties, as these will pull on the fits.  Unfortunately, there are no clear cuts in galaxy properties like inclination or observed size that will provide a cleaner sample.  Given this, we simply fit these gas-only scaling relations using the full sample of 148 galaxies.}

A relatively straightforward approach to fitting \textcolor{black}{a scaling relation with differing uncertainties in both parameters} is to use orthogonal distance regression (ODR) where the distance $\delta_{i,p}$ of a line to a data point normalized by the uncertainty is minimized. 
Explicitly, the orthogonal distance of a data point to another point is
\begin{equation}\label{Eq:NormDist}
    \delta_{i,p}^{2} =\left(\frac{X_{i}-X_{p}}{\epsilon_{X,i}}\right)^{2}+\left(\frac{Y_{i}-Y_{p}}{\epsilon_{Y,i}}\right)^{2}~,
\end{equation}
where $i$ denotes the data point, $p$ is a point on a line, and $\epsilon$ is the uncertainty in the $X$ or $Y$ coordinates respectively.  Note that when $X$ and $Y$ are logarithmic, the uncertainties must be in that same logarithmic space. 
For a given line, the total distance is the sum of $\delta_{i,p}$ where $p$ is explicitly the point of closest approach.  In ODR, the line that minimizes the total distance represents the best fit. 
In this work, we have adopted the ODR package implemented in \textsc{SciPy} \citep{2020SciPy-NMeth}.

A limitation of the \textsc{SciPy} ODR package is that it only allows for symmetric uncertainties \textcolor{black}{on a measured value (i.e.\ the upper and lower errorbars are equal)}. \textcolor{black}{Such a restriction is common in many tools, but it is important as, even when uncertainties are equal in linear space (which is not necessarily true for the WALLABY \textcolor{black}{\hi\ disk structural properties}), in the logarithmic space of scaling relations, such uncertainties are not equal.}  For simplicity, we present here the results of the ODR analysis using only the upper errors as they are slightly smaller than the lower errors on average.  However, the results are consistent regardless of which set of \textcolor{black}{uncertainties} are used for fitting.

In order to obtain the uncertainties on the best fit parameters a customized bootstrapping algorithm is adopted.  In a bootstrapped approach to uncertainties, the aim is to subtract a model, scramble the residuals, add the model back, and refit the data. Given that we are using ODR for the fitting, the residuals to be scrambled are the ODR distances.  Therefore, once the best fitting line is found we calculate the full set of residuals at their closest approach to the line.  Then, for each data point, $(X_{i},Y_{i})$, a random residual pair is drawn from the full set.  The bootstrapped point is then shifted to a new position $(X_{i}^{'},Y_{i}^{'})$ such that its \textcolor{black}{orthogonal} distance to the best fit line matches the drawn residual.  The uncertainties \textcolor{black}{in the scaling relations} are calculated by generating $500$ different bootstrapped samples and re-fitting each sample.  Then the uncertainty in either the slope or intercept of the best fitting line is set to
\begin{equation}\label{Eq:Uncertainty}
    \delta \gamma^{2}=\textrm{std}(\gamma_{bs})^{2}+\left(\overline{\gamma_{bs}} -\gamma_{bf}\right)^{2}~,
\end{equation}
where $\gamma$ is either the slope or intercept of the fit, $\gamma_{bf}$ is the best fitting value, $\textrm{std}(\gamma_{bs})$ is the standard deviation of the bootstrapped distribution and $\overline{\gamma_{bs}}$ is the mean of the bootstrapped sample. \textcolor{black}{In other words, the uncertainty in the fit is the sum of the standard deviation of the bootstrap fit distribution and the difference between the best fit value and the bootstrap mean in quadrature.}

\begin{figure*}
\centering
    \includegraphics[width=\textwidth]{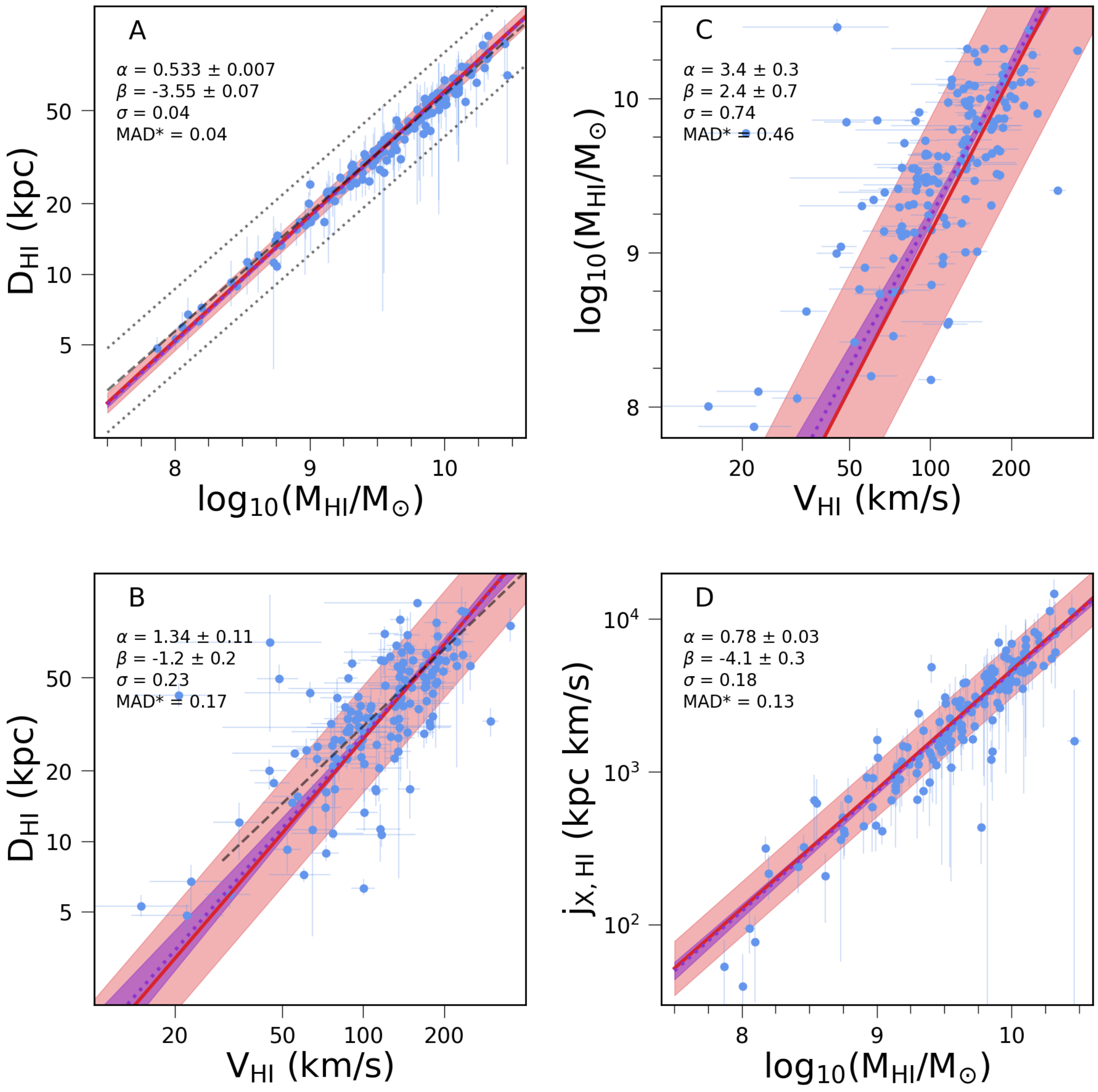} 
    \caption{The results of the bootstrap ODR analysis for the size-mass (A), size-velocity (B), \textcolor{black}{mass-velocity} (C), and angular momentum-mass (D) scaling relations.  The dashed grey lines in panels A and B are the \citet{Wang2016} size-mass and \citet{Meurer2018} size-velocity relations respectively, \textcolor{black}{while the dotted grey lines in panel A are the $\pm~3~\sigma$ lines from \citet{Wang2016})}.  The solid red lines are the best fit lines to the WALLABY data.  The shaded red region shows the 1 $\sigma$ \textcolor{black}{vertical} scatter of points about the best fit lines.  The \textcolor{black}{dark purple dashed line shows the running average of the 500 bootstrap samples, while the shaded region shows the running $1~\sigma$ width of the distribution.}}
  \label{Fig:BootstrapScalingRelations_Limited}
\end{figure*}

Figure \ref{Fig:BootstrapScalingRelations_Limited} shows the bootstrapped ODR fits graphically, while Table \ref{tab:ODR_Results} lists the best fit scaling parameters.  In addition to the \textcolor{black}{slope and intercept}, $\alpha$ and $\beta$, \textcolor{black}{we also calculated the standard deviation $\sigma$ and the median absolute derivation (MAD) of the observed vertical scatter from the best fit line.}  If the distribution is a Gaussian with no outliers, then the relation between the scatter and MAD is given by $\sigma=1.4826\rm{MAD}$ \citep{Rousseeuw1993}. For ease in comparison with standard scatter measures, in both Figure \ref{Fig:BootstrapScalingRelations_Limited} and Table \ref{tab:ODR_Results} we record $\rm{MAD*}=1.4826\rm{MAD}$. \textcolor{black}{In Figure \ref{Fig:BootstrapScalingRelations_Limited}, the shaded red regions correspond to $\sigma$}, while the \textcolor{black}{dark purple} dashed lines and shaded regions indicate the running average and widths  of the bootstrapped distribution respectively.  The \textcolor{black}{purple} lines and regions are analogous to (but not exactly the same as) $\overline{\gamma_{bs}}$ and $\textrm{std}(\gamma_{bs})$ in Eq. \ref{Eq:Uncertainty}.  The difference between the \textcolor{black}{purple} shaded region and red shaded region in each panel indicates the difference in range of allowable scaling relations compared to the width of the galaxy property distribution.

As expected, the size-mass relation is very tight with small scatter about the best fit line. \textcolor{black}{Comparing to the results from \citet{Wang2016}, the uncertainty in the intercept is approximately the same (\textcolor{black}{0.007 compared to 0.009} in \citealt{Wang2016}), while the uncertainty in the slope is larger (0.003 compared to our \textcolor{black}{0.007}). 
However, the vertical scatter of the galaxies about the best fit relation is smaller for the WALLABY sample (0.06 dex in \citet{Wang2016} compared to 0.04 dex here).} \textcolor{black}{Since the mean uncertainty in $D_{\hi}$ is $\sim$0.05 dex, it is plausible that all the scatter} \textcolor{black}{in this relation can be explained by observational uncertainties.  Confirming this will require a more advanced fitting method where the intrinsic scatter is also modeled.}

\begin{deluxetable*}{ c c c c c c c c c c }
\tablewidth{1.0\textwidth} 
\tablehead{
\colhead{Scaling Relation}       & \colhead{$X$}       &  \colhead{$Y$}       & \colhead{$\alpha$}       & \colhead{$\delta \alpha$}       & \colhead{$\beta$}       & \colhead{$\delta \beta$}       & \colhead{$\sigma$ [dex]}                                                 & \colhead{MAD* [dex]}
}
\startdata
        \hline 
        Size-Mass       & $\log_{10}(M_{\hi})$       & $\log_{10}(D_{\hi})$       & 0.533       & 0.007       & -3.55       & 0.07       & 0.04       & 0.04\\
        Size-Velocity       & $\log_{10}(V_{\hi})$       & $\log_{10}(D_{\hi})$       & 1.34       & 0.11       & -1.2       & 0.2       & 0.23       &0.17\\
        \textcolor{black}{mass-velocity}       & $\log_{10}(V_{\hi})$       & $\log_{10}(M_{\hi})$       & 3.4       & 0.3       & -2.4       & 0.7                                                & 0.74                                                &0.46\\
        Angular Momentum-Mass       & $\log_{10}(M_{\hi})$       & $\log_{10}(j_{\rm{X},\hi})$       & 0.78       & 0.03       & -4.1       & 0.3       & 0.18                                                &0.13\\
        \hline 
        Stellar Tully-Fisher       & $\log_{10}(V_{\hi})$       & $\log_{10}(M_{*})$       & 4.1       & 0.3       & 1.2       & 0.7       & 0.50       &0.31\\
        Baryonic Tully-Fisher       & $\log_{10}(V_{\hi})$       & $\log_{10}(M_{B})$       & 3.3       & 0.2       & 3.2       & 0.5       & 0.39       &0.22\\
        \hline
\enddata
\caption{The slope ($\alpha$), intercept ($\beta)$, and associated uncertainties for each of the scaling relations examined in this work. 
 These values are obtained by the ODR bootstrap method described in Sec. \ref{Sec:ScalingRelations}.  Finally, $\sigma$ is the standard deviation of the observed vertical scatter in dex from the best fit line.}
    \label{tab:ODR_Results}
\end{deluxetable*}

\textcolor{black}{In contrast to the size-mass relation}, the size-velocity relation shows a significant amount of scatter. However, the fit over the range $ 50\,\kms \lesssim V_{\hi} \lesssim 200~\kms$ agrees with the results from \citet{Meurer2018}.  Moreover, the observed size scatter is similar to that found by \citet{Meurer2018} ($\rm{MAD}*=0.16$ compared to their $\sigma_{R}=0.18$).

\textcolor{black}{Moving to the \textcolor{black}{mass-velocity} and angular momentum-mass relations in panels C and D of Figure \ref{Fig:BootstrapScalingRelations_Limited}, the scatter in the \textcolor{black}{mass-velocity} relation is relatively high, much of it is driven by low velocity outliers (as indicated by the difference between the $\sigma$ and $\rm{MAD}*$ statistics).  \textcolor{black}{One potential cause of the low velocity outliers are errors in the inclination measurement.  We have tested this possibility using the inclination matched subsample.  This removes many of the outliers, but not all.  As such, we conclude that inclinations are not the only source of low velocity outliers.  Regardless of the sample used, this} is one of the first quantitative measures of the \hi\ \textcolor{black}{mass-velocity} relation, which is a helpful counterpart to the sTFR and bTFR.  The angular momentum-\hi\ mass relation is also well constrained by our fit, which covers many orders of magnitude in both mass and angular momentum.  \textcolor{black}{Although the relation appears visually tight, the scatter is comparable to that of the size-velocity relation. }
}

\section{Stellar and Baryonic Tully-Fisher Relations}\label{Sec:bTFR}

\begin{figure*}
\centering
    \includegraphics[width=0.7\textwidth]{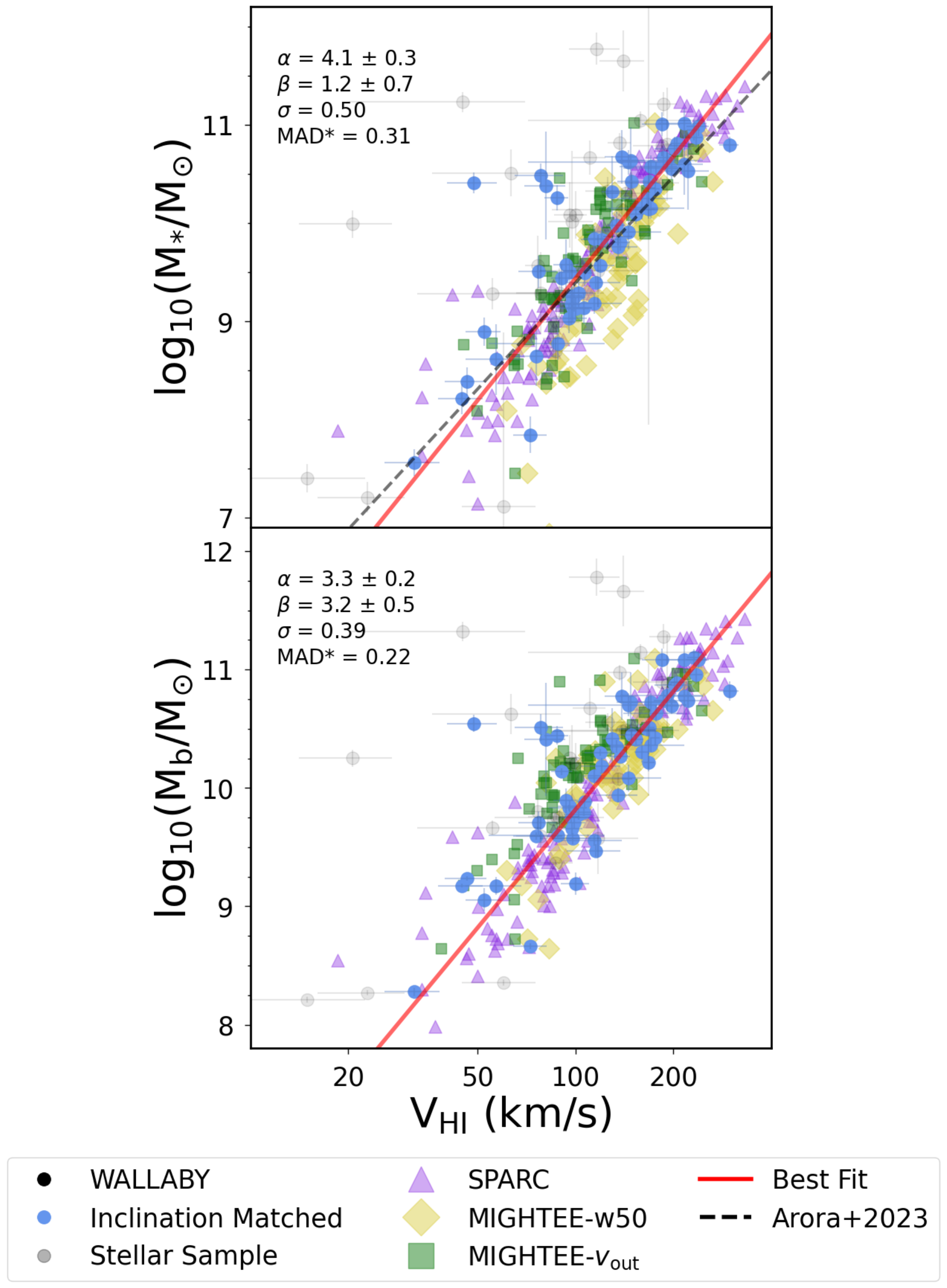} 
    \caption{The stellar (upper panel) and baryonic (lower panel) Tully-Fisher relation.  The solid red line is the best fit line from the ODR bootstrap analysis. The SPARC and MIGHTEE samples are obtained from \citet{Lelli2016} and \citet{Ponomareva2021} respectively, while the dashed black line is from the \citet{Arora2021} analysis of MaNGA data \citep{Bundy2015,Wake2017}.
    }
  \label{Fig:bTFR}
\end{figure*}

Equipped with the stellar masses determined in Sec. \ref{ssec:StellarMass}, we can examine the sTFR and bTFR relations for the WALLABY galaxies. 
\textcolor{black}{These are two of the most well studied galaxy scaling relations \citep{Tully1977,Mcgaugh2000, Verheijen2001, Hall2012, Lelli2016,Papastergis2016, Ponomareva2017,Lelli2019,Geha2016,McQuinn2022,Arora2023}, and are used to identify drivers of galaxy diversity as well as provide powerful constraints on cosmological simulations \citep{Trujjilo2011, Dutton2011, Brook2012, Arora2023}. }
\textcolor{black}{This analysis uses the stellar inclination matched sample of 61 galaxies in order to minimize effects from erroneous kinematic inclinations on $V_{\hi}$. Fits for both the sTFR and bTFR are obtained using the same bootstrapping approach of Sec. \ref{Sec:ScalingRelations}.  These fits are shown in Figure \ref{Fig:bTFR} and the results are listed in Table \ref{tab:ODR_Results}.}

The upper panel of Figure \ref{Fig:bTFR} shows the sTFR for the resolved WALLABY galaxies while the bottom panel shows the bTFR.  \textcolor{black}{The blue points show the inclination matched sample and grey points show the remaining 92 galaxies in the full stellar sample.  It is clear that those galaxies with discrepant inclinations tend to have lower velocities than expected from best fitting relations.  Nonetheless, there are some remaining outliers in the inclination matched sample.  The most likely origin for these is that the flat disk models generated by \wkapp\ are not appropriate for those particular galaxies, which will be investigated fully in future work.}

\textcolor{black}{Figure \ref{Fig:bTFR} also compares the WALLABY sample to the SPARC (purple triangles; \citealt{Lelli2016} and MIGHTEE-HI samples (yellow diamonds and green squares; \citealt{Ponomareva2021}) as well as the bTFR in \citet{Arora2023} derived from MaNGA data \citep{Bundy2015,Wake2017}.  The SPARC sample is described in Sec. \ref{Sec:GasOnlyRelations}.  MIGHTEE is the MeerKAT International GigaHertz Tiered Extragalactic Exploration survey \citep{Jarvis2016}, and MIGHTEE-HI \citep{Maddox2021} is the \hi\ focused component of that survey.  In \citet{Ponomareva2021} they constructed bTFR measures using both the inclination corrected profile widths and the outermost velocity, $V_{\rm{outer}}$ found in a \barolo\ analysis of the data. For completion Figure \ref{Fig:bTFR} shows both MIGHTEE-HI velocity measures.  \citet{Arora2023} examined 2368 galaxies from the Mapping Nearby Galaxies at APO  survey (MaNGA; \citealt{Bundy2015,Wake2017}), which is a large scale IFU survey.  Using the MaNGA detections, \citet{Arora2023} examined their dynamic properties and calculated their scaling relations, including their sTFR (the dashed grey line in the upper panel of Figure \ref{Fig:bTFR}).
}

\textcolor{black}{The WALLABY galaxies show a tight sTFR with a scatter that is comparable to the SPARC sample and both MIGHTEE-HI samples.  However, both measures of the WALLABY sTFR scatter, $\sigma=0.47$, and $\rm{MAD}*=0.35$, are significantly larger than $\sigma=0.24$ scatter found in \citet{Arora2023}.  It is possible that the increased scatter is due to the remaining outliers, but it may also be due to the WALLABY galaxies sampling different regions of parameter space with differing properties than \citet{Arora2023}.}

\textcolor{black}{
Qualitatively, the WALLABY bTFR is also generally consistent with both the SPARC and MIGHTEE-HI samples.  Interestingly, the MIGHTEE-HI inclination-corrected $w_{50}/2$ velocity measure (yellow diamonds) is more consistent with SPARC and WALLABY than their $V_{\rm{outer}}$ measure derived from their \barolo\ analysis.  As shown in Figure 3 of \citet{Ponomareva2021} the \barolo\ velocities are systematically lower than the corrected $w_{50}/2$ values, which is likely due to their \barolo\ measures likely probing a different portion of the RC (see also \citealt{Verheijen2001,Noordermeer2007,Ponomareva2018,Lelli2019}.  While \citet{Arora2023} also measure a bTFR we do not show it in the lower panel of Figure \ref{Fig:bTFR} as the MaNGA field of view is smaller than WALLABY and consequently, the \citet{Arora2023} gaseous and baryonic masses are lower than those derived from \hi\ surveys.}

\textcolor{black}{Quantitatively, \citet{Ponomareva2021} measure a scatter of $\sigma \sim0.32$ for their bTFR regardless of the velocity measure used.  This value is similar to the WALLABY $\sigma$, but the WALLABY scatter is largely driven by a few outliers.  Our $\rm{MAD}*$ bTFR statistic is 0.19, which, coupled with a visual inspection, suggests that the WALLABY sample has slightly less scatter than the MIGHTEE-HI sample.  However, a more in depth analysis will be needed to confirm this.  Additionally, \citet{Ponomareva2021} obtain an estimate of the intrinsic scatter, which we are unable to measure with our fitting current technique. Moving to SPARC, \citet{Lelli2019} use a different scatter parameterization that includes a measure of the instrinsic scatter, which makes a quantitative comparison between the scatter of the SPARC and WALLABY unfeasible for this work.  With full the WALLABY survey data we will be able to measure the intrinsic scatter, allowing for direct comparisons between SPARC, MIGHTEE-HI, and WALLABY. 
}

\textcolor{black}{Interestingly the slope of the \hi\ mass-velocity relation is consistent with the slope of the bTFR (with very different intercepts).  This result implies a linear relation between $M_{\hi}$ and $M_{B}$.  Such a relation may simply be a consequence of the WALLABY selection function, which preferentially detects gas-rich galaxies. }

\textcolor{black}{Following \citet{Lelli2019} and comparing the scatter of the \textcolor{black}{\hi\ mass-velocity relation}, the sTFR, and the bTFR, it is clear that the bTFR shows the least scatter.  This result, which is the same as \citet{Lelli2019}, suggests that, for full WALLABY, probes of the bTFR in different environments will be more likely to reveal differences in the scaling relation arising due to environment than the sTFR or \textcolor{black}{\hi\ mass-velocity relation}. }

\section{Gas Fraction and Angular Momentum}\label{Sec:QAnalysis}

\begin{figure*}
\centering
    \includegraphics[width=0.85\textwidth]{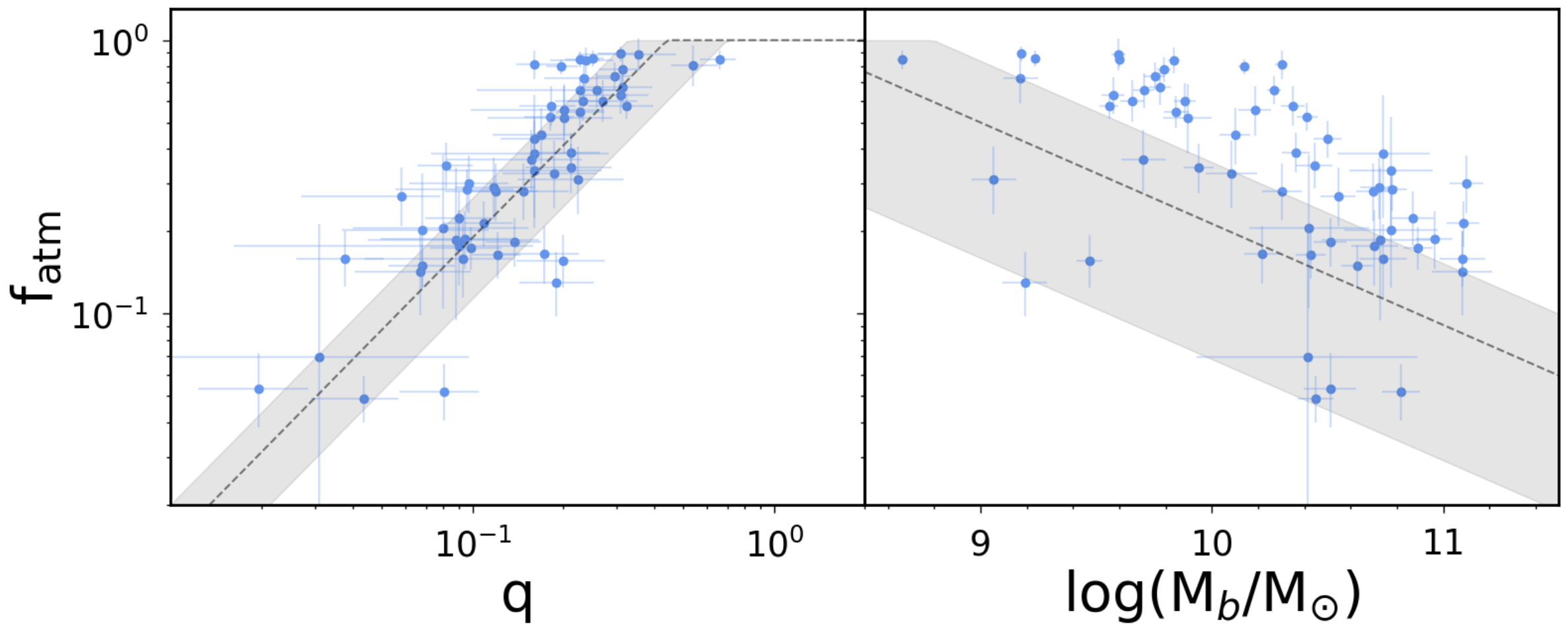} 
    \caption{The neutral atomic gas fraction relations to the disk stability (left panel) and baryonic mass (right panel).  In the left panel the dashed grey line shows Eq. \ref{Eq:GasFrac_Q}, while in the right panel the dashed grey line shows Eq. \ref{Eq:GasFrac_M}.  The shaded grey region on the left is the $\pm40\%$ region based on the expected scatter in $\sigma$ for galaxies \citep{Obreschkow2016}.  The shaded region in the right panel shows a $\pm68\%$ scatter that \citet{Obreschkow2016} predict for this relation based on the distribution of model halo spins \citep{Bullock2001}.
    }
  \label{Fig:QRelations}
\end{figure*}

\textcolor{black}{\citet{Obreschkow2016} posit from theoretical arguments that angular momentum regulates the gas fractions of galaxies. In their models of exponential disks, the gas fraction is related to the global disk stability via}
\begin{equation}\label{Eq:GasFrac_Q}
    f_{\rm{atm}}=\rm{min}\left[1,2.5q^{1.12}\right]~.
\end{equation}
\textcolor{black}{This gas fraction is also related to the baryonic mass of the galaxy via}
\begin{equation}\label{Eq:GasFrac_M}
    f_{\rm{atm}}\approx 0.5 \left(\frac{M_{b}}{10^{9}M_{\odot}}\right)^{-0.37}~.
\end{equation}
\textcolor{black}{Figure \ref{Fig:QRelations} shows a comparison of the inclination matched stellar sample to these two relations using $q_{\rm{X}}$ as a proxy for $q$.  However, $q_{\rm{X}}$ is essentially a maximal/outer stability parameter rather than the global stability parameter of \citet{Obreschkow2016}.}

The left-hand panel of Figure \ref{Fig:QRelations} shows that the WALLABY galaxies
have a slight offset towards lower values of $q_{\rm{X}}$ for a given gas fraction than predicted by Eq. \ref{Eq:GasFrac_Q}.  \textcolor{black}{\citet{Hardwick2022} also probed this relation using the eXtended GALEX Arecibo SDSS Survey \citep{Catinella2010,Catinella2018}.  They find a similar offset over the range of $0.1 \le f_{\rm{atm}}\le 1.0$, but, at lower gas fractions, they find a different slope and a larger scatter than predicted by \citet{Obreschkow2016}.  In the range of WALLABY gas fractions, the majority of galaxies in both the WALLABY and \citet{Hardwick2022} samples lie within the $\pm40\%$ region suggested by \citet{Obreschkow2016} and have a similar slope as the predicted line}.  This result is slightly different than the analysis of WALLABY pre-pilot observations in Eridanus by \citet{Murugeshan2021}, who found an offset towards higher $q_{\rm{X}}$ values.  A key difference is that \citet{Murugeshan2021} were able to extract a measure of $q$, as opposed to our use of $q_{\rm{X}}$.  For future studies of the disk stability we will need to either reduce the sample further to those where $q$ can be reliably calculated or return to the \citet{Obreschkow2016} formalism and find a theoretical model of $q_{\rm{X}}$.

\textcolor{black}{Moving to the gas fraction-baryonic mass relation, the WALLABY galaxies tend to lie above the predicted relation given by Eq. \ref{Eq:GasFrac_M}.  This is similar to what is seen in \citet{Obreschkow2016} for most galaxies below $10^{10}~M_{\odot}$, where their sample also lies above the relation.  At higher masses, their sample becomes more consistent with the prediction, which may be true for WALLABY, but \textcolor{black}{a larger sample of high mass kinematic} models will be needed to fully confirm this behaviour.  \textcolor{black}{It is likely that the bias towards higher $f_{\rm{atm}}$ at a given baryonic mass is due to WALLABY being an HI selected survey, which is naturally biased towards gas rich galaxies.  Testing this possibility will require a larger sample spanning a range of gas richness.}  \citet{ManceraPina2021b} also examined \textcolor{black}{gas fraction-baryonic mass relation} in the context of a $j_{b}-M_{b}-f_{\rm{atm}}$ plane. They \textcolor{black}{also see that the gas fraction decreases with baryonic mass} and predict that the ratio of the gaseous to stellar specific angular momentum in these galaxies controls the scatter seen in the right-hand panel of Figure \ref{Fig:QRelations}.  \textcolor{black}{Similarly, \citet{Hardwick2023} examined these relations and find that xGASS observations are consistent with a variety of cosmological simulations, arguing that the \citet{Obreschkow2016} model needs to be expanded to fully capture the physics of angular momentum and gas fractions.  With the full WALLABY survey, coupled with multiwavelength observations, it will be possible to confirm these predictions.
}}

\section{Discussion and Conclusions}\label{Sec:Conclusions}

The kinematically modelled WALLABY PDR galaxies are currently the largest sample of untargetted, uniformly observed and analyzed \hi\ kinematic models available.  \textcolor{black}{This makes them the ideal sample for studies of scaling relations that involve \hi.  In this work we have designed a framework for calculating \textcolor{black}{\hi\ disk structural properties} that is applicable to the marginally and moderately resolved WALLABY galaxies, that is robust and scalable for the full WALLABY sample. Armed with robust \textcolor{black}{structural properties}, we have obtained the first measures of the size-mass, size-velocity, \textcolor{black}{mass-velocity}, angular momentum-mass, sTFR, bTFR, and gas fraction scaling relations for the WALLABY survey.}

\textcolor{black}{The key \textcolor{black}{\hi\ disk structural properties} we attempt to extract are the disk size, outer velocity, and angular momentum.  We measure the disk size, $R_{\hi}$, as the radius where $\Sigma(R_{\hi})=1~\Msol~\rm{pc}^{-2}$.  We characterize the disk velocity as $V_{\hi}=V(R_{\hi})$ rather than $v_{\rm{flat}}$ in order to include galaxies with potentially rising or falling rotation curves.  And, due to beam-smearing effects in the integration of the specific angular momentum, we characterize the angular momentum as $j_{\rm{X},\hi}=1/2 D_{\hi}V_{\hi}$.  For those with available optical DESI 3-band images, we also extract the stellar mass, the baryonic mass, the atomic gas fraction, and the disk stability, $q_{\rm{X}}=j_{\rm{X},\hi}\sigma/(G M_{\rm{bary}})$.}

\textcolor{black}{While there are 236 available WALLABY kinematic models, many of these are marginally resolved and do not have robustly measured \textcolor{black}{structural properties}.  However, the advantage of such a large sample is the ability to curate it and probe scaling relations using only the well measured galaxies.  There are 148 galaxies with robustly measured sizes, velocities, and angular momenta. }

\textcolor{black}{The WALLABY galaxies match the \hi\ size-mass relation of \citet{Wang2016}, with a very tight scatter about the relation. This scatter, which can be fully explained by observational uncertainties, may actually be smaller than the \citet{Wang2016} sample, although it spans a much smaller range of masses.  The WALLABY galaxies also match the \citet{Meurer2018} size-velocity relation, albeit with much more scatter than the size-mass relation.  However, the WALLABY sample size is much larger than the \citet{Meurer2018}, allowing for a much more robust measure of this scaling relation.  Moving to the other relations, we have obtained one of the first measures of the \hi\ \textcolor{black}{mass-velocity} relation.  A clear relation is seen, albeit with a great deal of scatter.  And, while there are other empirical relations probing the specific angular momentum-mass relation, our adoption of $j_{\rm{X},\hi}$ prevents a quantitative comparison of those relations to our results.}

\textcolor{black}{Beyond just empirical relations, we also compared the WALLABY sample to SPARC galaxies \citep{Lelli2016} and LVHIS galaxies \citep{Koribalski2018}.  In general the three samples agree for all four of these \hi\ scaling relations.   }

\textcolor{black}{For studying the sTFR and bTFR, as well as the disk stability and gas fraction, we use a slightly different sample that leverages the power of our optical measurements.  There are 92 galaxies with robust \hi\ \textcolor{black}{structural properties} and stellar mass measurements.  The further constraint that the optical and kinematic angles are consistent reduces the sample to 61 galaxies.  And it is this inclination matched sample that we use to the constrain the sTFR and bTFR. }

\textcolor{black}{The WALLABY sTFR is consistent with the sTFR measured by \citet{Arora2023} using 2368 MaNGA galaxies.  It is also consistent with SPARC and MIGHTEE-HI measures of the sTFR.}
The WALLABY bTFR shows less scatter than the sTFR, which is consistent with \citet{Lelli2019}.  This suggests that the bTFR is a more fundamental relation than the sTFR \textcolor{black}{(as previously stated in \citet{Mcgaugh2000}) and \citet{Lelli2019}}.  \textcolor{black}{There are some outliers remaining in our inclination matched sample that will be investigated in future work. Additionally, as WALLABY continues we expect to reach to lower baryonic masses and enabling deeper investigations of the TFRs.}

Comparing the WALLABY galaxies gas fraction scaling relations to the theoretical models of \citet{Obreschkow2016} show some interesting behaviour.  As expected, the gas fraction increases with the disk stability parameter and decreases with the baryonic mass.  The WALLABY galaxies are offset from the \citet{Obreschkow2016} disk stability relation, but this is likely due to our use of $q_{\rm{X}}$ rather than $q$.  There is also a \textcolor{black}{significant} offset from the gas fraction-baryonic mass relation, \textcolor{black}{likely due to the untargetted nature of WALLABY preferentially detecting galaxies with high gas fractions.}

\textcolor{black}{Moving forwards, there are two key limitations of this study that will need to be addressed.  Firstly, it will be necessary to extend the concept of orthogonal distances to include fully asymmetric uncertainties in order to calculate a more robust likelihood.  Secondly, it will be necessary to switch to Bayesian techniques that can fit the intrinsic scatter as separate parameters from the observed scatter.}

\textcolor{black}{WALLABY is the premier survey for studying gaseous scaling relations in the local Universe. The untargetted nature of the survey means that the selection effects and biases are easily understood.  And the wide area of the survey means that the number of detected and modelled galaxies will be far beyond any other contemporary survey.  The pilot data releases comprise approximately 1\% of the total survey area, and has already provided the largest sample of uniformly observed and analyzed kinematic models. This study is the first to study scaling relations using WALLABY kinematic models and provides the framework for future studies.  }  With the full set of $\sim10^{4}$ kinematic models expected for the WALLABY \citep{Koribalski2018,Westmeier2022}, we will improve the constraints beyond even this study, probe the intrinsic scatter of the various relations, and be able to investigate differences in scaling relations due to differing environments.

\section*{Data Availability}

ALL WALLABY PDR1 data is publicly available at \href{https://wallaby-survey.org/data/data-pilot-survey-dr1/}{WALLABY PDR1}.  
The kinematic modelling proto-pipeline is available at \href{https://github.com/CIRADA-Tools/WKAPP}{\wkapp\ code}.  
\textcolor{black}{A machine readable table containing all calculated \textcolor{black}{\hi\ disk structural properties} is available in the online Journal}.
The specific analysis scripts are available upon request.

\section*{Acknowledgements}

\textcolor{black}{We would like to thank the anonymous referee for their helpful comments and suggestions.}  \textcolor{black}{Special thanks to A. Ponomareva for helpful discussions on MIGHTEE-HI.  Thanks to Y. Ascasibar, A. Boselli, C. Carignan, E. Di Teodoro, J. English, B. Holwerda, and J.M. van der Hulst for helpful comments. KS acknowledges financial support from the Natural Sciences and Engineering Research Council of Canada (NSERC). AB acknowledges support from the Centre National d’Etudes Spatiales (CNES), France. LC acknowledges support from the Australian Research Council Discovery Project funding scheme (DP210100337). PK is partially supported by the BMBF project 05A23PC1 for D-MeerKAT. PEMP acknowledges the support from the Dutch Research Council (NWO) through the Veni grant VI.Veni.222.364.  Parts of this research were supported by the Australian Research Council Centre of Excellence for All Sky Astrophysics in 3 Dimensions (ASTRO 3D), through project number CE170100013.
}

\textcolor{black}{
This scientific work uses data obtained from Inyarrimanha Ilgari Bundara / the Murchison Radio-astronomy Observatory. We acknowledge the Wajarri Yamaji People as the Traditional Owners and native title holders of the Observatory site. CSIRO’s ASKAP radio telescope is part of the Australia Telescope National Facility (https://ror.org/05qajvd42). Operation of ASKAP is funded by the Australian Government with support from the National Collaborative Research Infrastructure Strategy. ASKAP uses the resources of the Pawsey Supercomputing Research Centre. Establishment of ASKAP, Inyarrimanha Ilgari Bundara, the CSIRO Murchison Radio-astronomy Observatory and the Pawsey Supercomputing Research Centre are initiatives of the Australian Government, with support from the Government of Western Australia and the Science and Industry Endowment Fund.  \textcolor{black}{WALLABY acknowledges technical support from the Australian SKA Regional Centre (AusSRC)
}}

\bibliography{new.ms}{}
\bibliographystyle{aasjournal}

\label{lastpage}
\end{document}